\documentclass[10pt,twocolumn]{revtex4-1}
\usepackage{lmodern}
\usepackage{lmodern}
\usepackage[latin9]{inputenc}
\usepackage{color}
\usepackage{amsmath}
\usepackage{amssymb}
\usepackage{graphicx}

\makeatletter
\usepackage{chemarr}

\makeatother

\begin{document}

\title{Grand canonical diffusion-influenced reactions: a stochastic theory
with applications to multiscale reaction-diffusion simulations}

\author{Mauricio J. del Razo$^{\dagger,a)}$, Hong Qian$^{\text{\ddag}}$
and Frank Noé$^{\dagger,a)}$}
\begin{abstract}
\textbf{Abstract:}Smoluchowski-type
models for diffusion-influenced reactions ($A+B\rightarrow C$) can
be formulated within two frameworks: the probabilistic-based approach
for a pair $A,B$ of reacting particles; and the concentration-based
approach for systems in contact with a bath that generates a concentration
gradient of $B$ particles that interact with $A$. Although these
two approaches are mathematically similar, it is not straightforward
to establish a precise mathematical relationship between them. Determining
this relationship is essential to derive particle-based numerical
methods that are quantitatively consistent with bulk concentration
dynamics. In this work, we determine the relationship between the
two approaches by introducing the grand canonical Smoluchowski master
equation (GC-SME), which consists of a continuous-time Markov chain
that models an arbitrary number of $B$ particles, each one of them
following Smoluchowski's probabilistic dynamics. We show the GC-SME
recovers the concentration-based approach by taking either the hydrodynamic
or the large copy number limit.
In addition, we show the GC-SME provides a clear statistical mechanical
interpretation of the concentration-based approach and yields an emergent
chemical potential for nonequilibrium, spatially inhomogeneous reaction
processes. We further exploit the GC-SME robust framework to accurately
derive multiscale/hybrid numerical methods that couple particle-based
reaction-diffusion simulations with bulk concentration descriptions,
as described in detail through two computational implementations.

\end{abstract}


\maketitle
\noindent $^{\dagger}$Freie Universität Berlin, Department of Mathematics
and Computer Science, Arnimallee 6, 14195 Berlin, Germany

\noindent $^{\ddagger}$Department of Applied Mathematics, University
of Washington, Seattle, WA 98195-3925.

\noindent $^{a)}$Corresponding authors. E-mails: \texttt{maojrs@gmail.com
and frank.noe@fu-berlin.de}

\noindent 

\section{Introduction \label{sec:intro}}

Smoluchowski's original diffusion-controlled reaction theory describes
the bimolecular reaction $A+B\to C$ in which diffusion is the transport
process in solution that determines the encounter between reacting
pairs \citep{collins1949diffusion,smoluchowski1917versuch}. In such
systems, the macroscopic bimolecular reaction rate depends on the
diffusion coefficients $(D_{A}+D_{B})$. This has
been an extremely successful model in physical chemistry \citep{agmon1990theory,hanggi1990reaction,keizer1982nonequilibrium,keizer1987diffusion,schurr-1970,szabo1980first,szabo1989theory},
with current on-going applications \citep{del2014fluorescence,del2016discrete,dorsaz2010diffusion,donev2010first,gopich2002kinetics,hagen1996diffusion,peters2013reaction,prustel2013theory,tucci2004mesoscopic,vijaykumar2015combining,vijaykumar2017multiscale}.
The theory assumes that a macromolecule $A$ sits at the origin surrounded
by a concentration gradient of $B$ molecules (ligands). If a $B$
molecule gets close enough to $A$, a reaction occurs and the $B$
is absorbed. This process is mathematically described by a diffusion
equation for the concentration gradient of $B$ with an absorbing
boundary condition around the $A$ molecule. 

A probabilistic version of the same model arose by interpreting Smoluchowski's
diffusion equation as a Fokker-Planck equation for one $B$ molecule
\citep{agmon1990theory,del2016discrete,szabo1980first,van2005green},
where the dynamics are now in terms of the \textit{probability density}
of finding $B$ at a certain point in space. The two models are mathematically
identical except for the far-field boundary condition.
The concentration-based approach assumes a constant concentration
at infinity, where else the probabilistic approach requires vanishing
density at infinity. As there is not yet a clear probabilistic interpretation
of the concentration-based approach, the two approaches seem incompatible
in a rigorous probability theory. Consequently, there is no theoretical
framework to develop probabilistic particle-based simulations that
are statistically consistent with bulk concentration dynamics, a highly
relevant issue for multiscale/hybrid reaction-diffusion simulations. This
brings to light the main key questions addressed in this work:
\begin{itemize}
\item What is the connection between Smoluchowski's probabilistic
and concentration-based approach? 
\item Can Smoluchowski's concentration-based approach
be interpreted in terms of a probabilistic model?
\item How can this connection be employed to develop particle-based
simulations that are consistent/coupled with bulk concentration descriptions?
\end{itemize}
These questions have been partly pointed out before \citep{szabo2008autobiography}
and have been somewhat solved \citep{agmon1990theory,berg1978diffusion,szabo2008autobiography}.
In the work \citep{agmon1990theory}, a microscopic theory (probabilistic)
of the kinetics of irreversible (and reversible) diffusion-influenced
reactions is developed and extended to pseudo-first-order reactions
(very large copy number of $B$s). In the thermodynamic
limit, this theory recovers the law of mass action with the corresponding
rate. However, the spatial information is lost and consequently the
full connection with Smoluchowski's concentration-based approach too.
More recent approaches developed in \citep{gopich2002asymptotic,gopich2002kinetics}
present a general theory of the kinetics of reversible diffusion-influenced
reactions. Unfortunately, it does not reduce to the Smoluchowski's
result in the irreversible limit. 

On the computational side, \citep{franz2013multiscale}
offers a good review of previous methods and presents a hybrid approach
to couple Brownian dynamics (particle-based) with reaction-diffusion
partial differential equations (PDEs) (concentration description).
However, it does not generalize to bimolecular reactions, and it is
only presented in one dimension, limiting its applicability.
Furthermore, \citep{smith2018spatial} offers an extensive review
that discusses the relationship between several particle-based and
master equation approaches, which complements many aspects of our
work.

In this work, we answer the three questions above
by developing a full stochastic theory of diffusion-influenced reactions,
called Smoluchowski Master equations (SMEs). Instead of using stochastic
diffusion processes in continuous space, SMEs are based on continuous-time
Markov chains, where the discrete state space simplifies the calculations
and lends itself to computational implementations.
Note that, although our work is based on Smoluchowski-type approaches,
the results are easily extendable to other diffusion-influenced reaction
models, like the Doi model \citep{doi1976stochastic}, which has certain
modeling and simulation advantages \citep{erban2009stochastic,isaacson2009reaction,isaacson2013convergent,smith2018spatial}.

We begin Sec. \ref{sec:modelsdiffinf} with an overview
of relevant diffusion-influenced reaction models.
Based on the probabilistic approach, we derive the first SME for an
isolated $A-B$ pair following \citep{del2016discrete}. In Sec. \ref{sec:grand},
we generalize the SME to an arbitrary nonconstant number of $B$ particles
by constructing the grand canonical Smoluchowski Master equation (GC-SME).
We further show Smoluchowski's concentration-based approach should
be understood as either the hydrodynamic limit (mean-field) or the
large copy number limit (law of large numbers) \citep{franco2014interacting,kipnis2013scaling}
of the GC-SME, a situation analogous to the Kurtz limit \citep{kurtz1971limit,kurtz1972relationship},
where the mass-action ordinary differential equation (ODE) is obtained
as the hydrodynamic/large copy number limit of the chemical master
equation \citep{qian2010chemical}. This result bridges the probabilistic
approach and the concentration-based approach, providing an unequivocal
interpretation of Smoluchowski's concentration-based approach in terms
of a probabilistic model. It further provides a statistical mechanical
interpretation of the concentration-based approach, which elucidates
interpretations at the particle level, and it establishes a connection
to nonequilibrium thermodynamics through the chemical potential. 

Secs. \ref{sec:app_nummeth} and \ref{sec:sim_exocyt}
show how the GC-SME framework can be employed to consistently couple
particle-based reaction-diffusion (PBRD) simulations with bulk concentration
dynamics in three dimensions. These multiscale/hybrid
schemes have several potential applications, such as: modeling of ion
channels \citep{corry2000tests} and exocytosis \citep{pedersen2011mathematical},
where single-particle resolution is fundamental in regions of relevance
but unnecessary in the far-field; and modeling of filopodial dynamics
\citep{zhuravlev2009molecular}, where the filopodia is in contact
with a larger cytosol compartment that can be modeled in the bulk.
These schemes are illustrated through two PBRD simulations that are
coupled to material baths: the first one emulates a constant concentration
bath, and the second one emulates a time- and space-dependent material
bath inspired by the exocytosis process from cell biology. The latter
is trivial to generalize to bulk concentration dynamics given in terms
of reaction-diffusion PDEs restricted to first-order reactions.

This work is a continuation of \citep{heuett2006grand} into spatiotemporal
stochastic chemical kinetics. The previous work provides a general
stochastic framework for open nonequilibrium linear networks constructed
through Markov chain theory.

\section{Models of diffusion-influenced reactions\label{sec:modelsdiffinf}}

In this section, we will review some of the main
models of diffusion-influenced reactions; more comprehensive descriptions
can be found in \citep{agmon1990theory,doi1976stochastic,hanggi1990reaction,schurr-1970,szabo1980first,szabo1989theory}.
Secs. \ref{sec:difreac} and \ref{sec:difreat-prob} show an overview
of the concentration-based and probabilistic approaches, as well as
how they differ from each other. Sec. \ref{sec:SME} derives the SME
for an isolated pair by discretizing the state space, which will serve
as an introduction to the GC-SME from Sec. \ref{sec:grand}.

\begin{figure*}[t]
	\centering \textbf{a.} \includegraphics[width=0.29\textwidth]{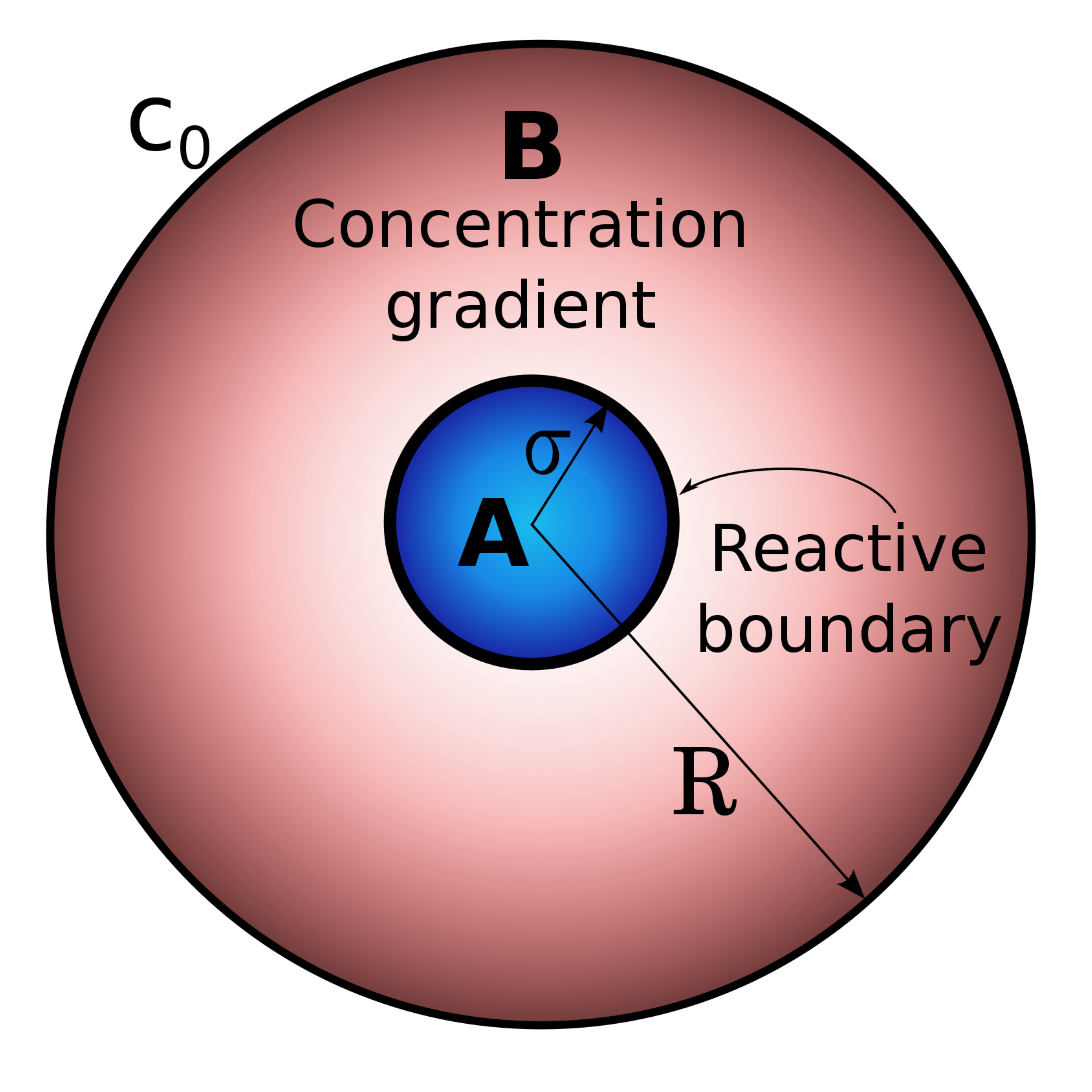}
	\textbf{b.} \includegraphics[width=0.29\textwidth]{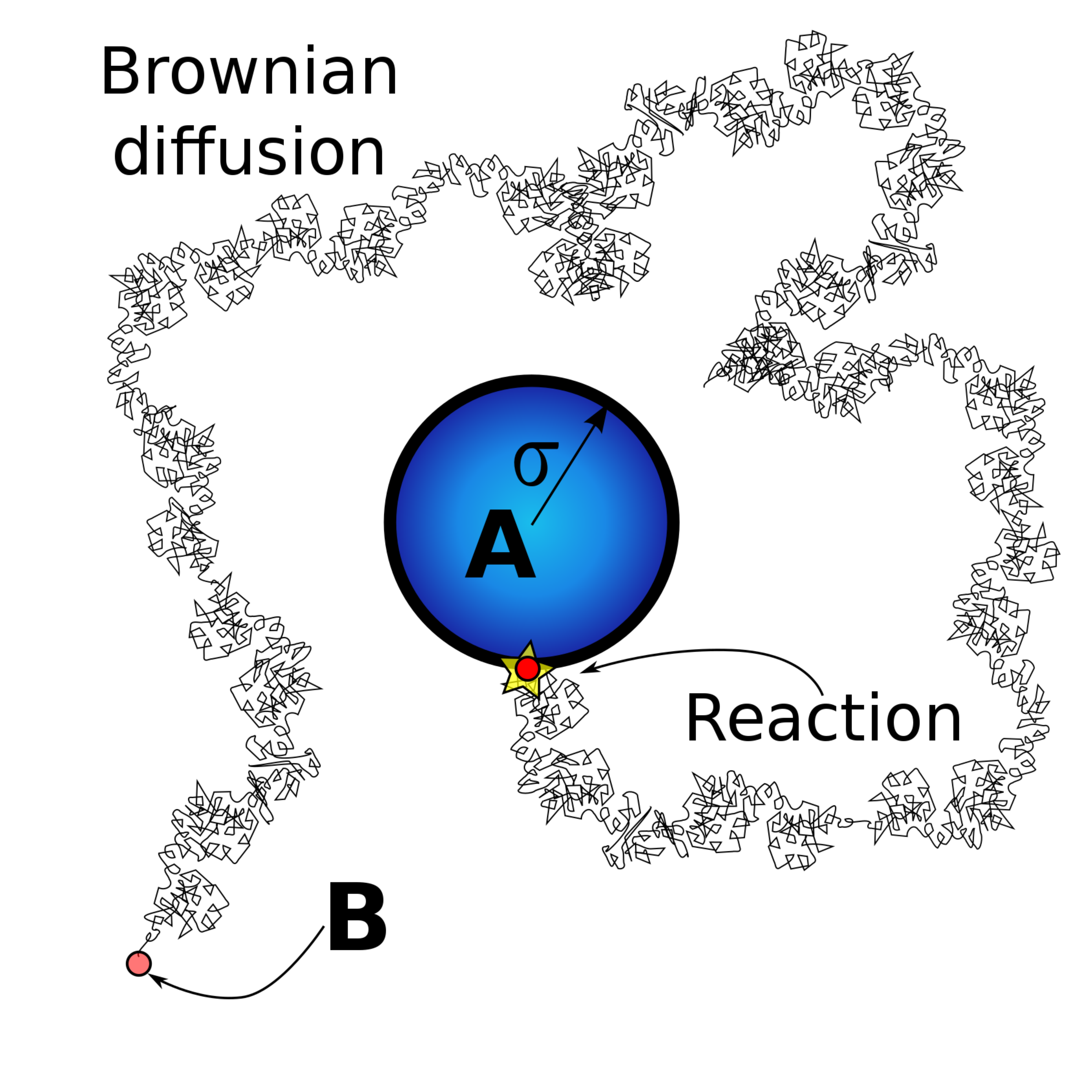}
	\textbf{c.} \includegraphics[width=0.29\textwidth]{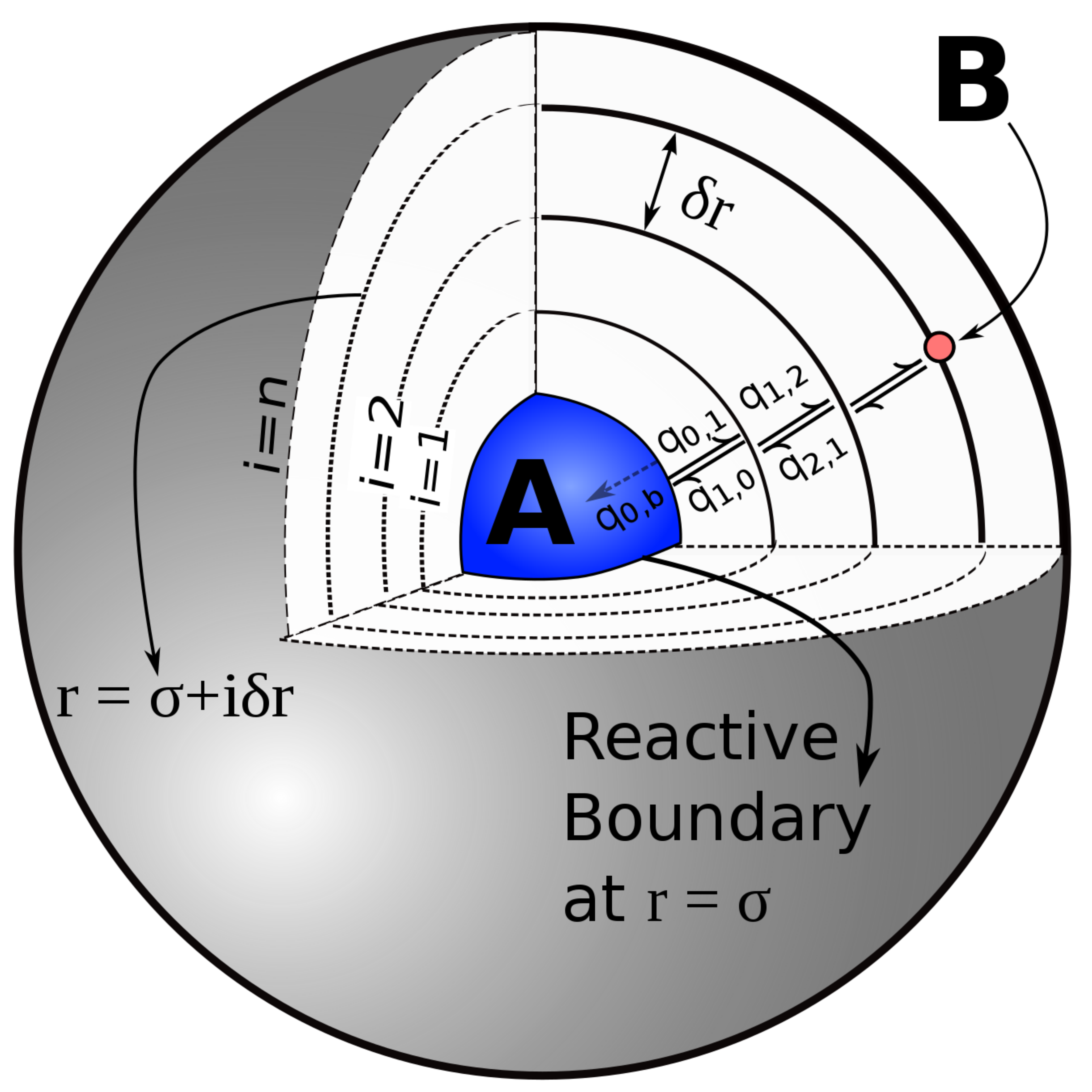}
	\caption{Illustrations of different models for diffusion-influenced reactions.
		In the three approaches, $A$ is fixed at the origin and $r=\sigma$
		represents the reaction boundary. \textbf{a.} The concentration-based
		approach: $A$ is surrounded by a concentration gradient of $B$ generated
		by a material bath with concentration $c_{0}$. \textbf{b.} The probabilistic
		approach: one $B$ diffuses around $A$ undergoing Brownian motion.
		\textbf{c.} The SME for an isolated pair: spatial discretization in
		spherical shells around $A$ of the probabilistic approach. Here we
		only track on which shell is the $B$ molecule, and the dynamics follow
		a continuous time Markov chain. }
	\label{fig:Approaches_cartoon} 
\end{figure*}

\subsection{Concentration-based approach \label{sec:difreac} }

The original diffusion-influenced reactions models
for bimolecular reactions $A+B\overset{}{\longrightarrow}C$ follow
a concentration-based approach \citep{collins1949diffusion,keizer1987diffusion,shoup1982role,smoluchowski1917versuch}.
In order to derive this model, we assume there is one $A$ represented
by a reactive sphere diffusing in space. The frame of reference is
fixed at the center of $A$, and $B$ molecules diffuse freely with
a diffusion coefficient given by $D=D_{A}+D_{B}$. The concentration
gradient of $B$ molecules around $A$ is denoted by $c(r,t)$, and
it obeys 

\begin{equation}
\frac{\partial c(r,t)}{\partial t}=\nabla\cdot\big[D\nabla c(r,t)\big].\label{eq:smol}
\end{equation}
The concentration in the far-field, $r=R$, is assumed constant, so
$c(R,t)=c_{0}$. The reaction is modeled by a reaction boundary $r=\sigma$
given by the sum of the molecules' radii. Whenever a $B$ molecule
reaches $\sigma$ by diffusion, a reaction occurs immediately. We
call this a purely diffusion-controlled reaction since the rate only
depends on the time it takes $B$ to diffuse into the reaction boundary.
This is modeled with a purely-absorbing boundary condition $c(\sigma,t)=0$,
which yields the steady state and the forward reaction rate 
\begin{equation}
c^{ss}(r)=c_{0}\left(\frac{R}{R-\sigma}\right)\left[1-\frac{\sigma}{r}\right],\ \ \ k_{S}(R)=4\pi D\sigma\left(\frac{R}{R-\sigma}\right)\label{eq:smol-pureabs}
\end{equation}
since $k_{S}(R)=4\pi D\sigma^{2}c^{ss'}(\sigma)/c_{0}$. As $R\rightarrow\infty$,
this simply becomes $c_{\infty}^{ss}(r)=c_{0}\left[1-\sigma/r\right]$
and $k_{S}=4\pi D\sigma$, which is Smoluchowski's original result
\citep{smoluchowski1917versuch}. A more general approach uses a partially-absorbing
boundary condition \citep{collins1949diffusion,shoup1982role}, $\left.4\pi\sigma^{2}D\partial c(r,t)/\partial r\right|_{r=\sigma}=\kappa c(\sigma,t)$,
where $\kappa$ controls the degree of diffusion influence in the
reaction rate. In this case, the steady state and reaction rate as
$R\rightarrow\infty$ are 
\begin{equation}
c^{ss}(r)=c_{0}\left[1-\frac{\kappa\sigma}{k_{S}+\kappa}\left(\frac{1}{r}\right)\right],\ \ \ \ \ k_{f}=\frac{\kappa k_{S}}{\kappa+k_{S}}.\label{eq:cksol}
\end{equation}
The purely diffusion-controlled result is recovered as a special case
in the limit $\kappa\rightarrow\infty$. 

\subsection{Probabilistic approach (for isolated pairs) \label{sec:difreat-prob} }

If we consider an isolated pair, one $A$ and one $B$, we can derive
a probabilistic theory for diffusion-influenced reactions \citep{agmon1990theory,sokolowski2010green,van2005green,van2005simulating,szabo1980first}.
Consider $A$ is fixed in the origin and $B$ diffuses following standard
Brownian motion. We denote $f(r,t|r_{0})$ the probability of molecule
$B$ being a distance $r$ from $A$ at time $t$ given that it was
at $r_{0}$ at time $0$. This transition probability will obey the
Fokker-Planck equation 
\begin{align}
\frac{\partial f(r,t|r_{0})}{\partial t} & =\nabla\cdot\big[D\nabla f(r,t|r_{0})\big],\label{eq:smol2}\\[3mm]
f(r,0|r_{0}) & =\frac{\delta(r-r_{0})}{4\pi r_{0}^{2}},\label{eq:smol2IC}\\
\left.4\pi\sigma^{2}D\frac{\partial f(r,t|r_{0})}{\partial r}\right|_{r=\sigma} & =\kappa f(\sigma,t|r_{0}),\label{eq:smol2BC}\\
\\
\lim_{r\rightarrow\infty}f(r,t|r_{0}) & =0.\label{eq:smol2FF}
\end{align}
Eq. (\ref{eq:smol2IC}) is the initial condition for the $B$ molecule;
Eq. (\ref{eq:smol2BC}) models the reaction boundary, and Eq. (\ref{eq:smol2FF})
corresponds to vanishing the probability as $r\rightarrow\infty$
due to normalizable total probability. Note this is a well defined
stochastic process, where $f(r,t|r_{0})$ is the ``remaining'' probability
density function in the presence of an absorbing boundary (diffusion
with killing). 

Although mathematically similar, the probabilistic approach from Eq.
(\ref{eq:smol2}) seems to be somewhat incompatible with the concentration-based
approach from Eq. (\ref{eq:smol}). This is due to the difference
in the outer boundary condition at $r=R$ (or $r=\infty$) and to
dealing with a probability instead of a concentration. In Sec. \ref{sec:grand},
we will understand how these two approaches are related, and the advantages
of understanding this relationship.

Note there are alternative models of the reaction
process, such as the purely absorbing reaction boundary, $f(\sigma,t|r_{0})=0$,
or even a volume reactivity model, like the Doi model \citep{doi1976stochastic},
which has been recently unified with Smoluchowski-type approaches
\citep{szabo1980first,ye2018dynamic}. Regardless of the reaction
process, the main results in Sec. \ref{sec:grand} and the computational
schemes from Secs. \ref{sec:app_nummeth} and \ref{sec:sim_exocyt}
remain valid since the results mainly concern the far-field boundary.

\subsection{Smoluchowski master equation \label{sec:SME} }

In \citep{del2016discrete}, a discrete time and state Markov chain
model for diffusion-influenced reactions for isolated pairs was derived.
This model recovers the probabilistic model from Sec. \ref{sec:difreat-prob}
in the continuous space limit. In this section, we will rewrite this
model as a Master equation (continuous-time Markov chain). 

Consider a macromolecule $A$ fixed at the origin and a ligand $B$
diffusing in the space around it. As we are interested in the diffusive
jumps of $B$ in the $r$ direction, we partition the space around
$A$ in spherical shells of width $\delta_{r}$ (Fig. \ref{fig:Approaches_cartoon}c).
If the particle is in shell $i$ with radius $r_{i}=\sigma+i\delta r$,
the probabilities to jump to the smaller and bigger shells are $\tilde{q}_{i,i-1}$
and $\tilde{q}_{i,i+1}$, given by $\tilde{q}_{i,i\pm1}=\delta t\left(D/\delta r^{2}\pm D/(r_{i\pm1}\delta r)\right)$.
This process is a discrete-time Markov chain $\boldsymbol{\pi}^{t+1}=\boldsymbol{\pi}^{t}\mathbb{P}$,
where $\pi_{i}^{t}$ is the probability of $B$ being at spherical
shell $i$ at time $t$, $\boldsymbol{\pi}^{t}=\left[\pi_{0}^{t},\pi_{1}^{t},\dotsc,\pi_{i}^{t},\dotsc\right]$,
and $\mathbb{P}$ the stochastic matrix in terms of $\tilde{q}_{i,i\pm1}$.
The probability of a reaction $A+B\rightarrow C$ is incorporated
into $\mathbb{P}$ by adding $\tilde{q}_{0,b}=\tilde{\kappa}(r)\delta t$,
at the innermost shell $r_{0}=\sigma$ (reaction boundary)\citep{del2016discrete}.
In order to obtain a Master equation, we can subtract $\boldsymbol{\pi}^{t}$
on both sides, divide by $\delta t$ and take the limit $\delta t\to0$
to obtain the SME 
\begin{equation}
\frac{d\boldsymbol{\pi}(t)}{dt}=\boldsymbol{\pi}(t)\mathbb{Q},\label{eq:SME}
\end{equation}
where $\boldsymbol{\pi}(t)$ is the continuous time analog of $\boldsymbol{\pi}^{t}$
and the matrix $\mathbb{Q}$ is given by

{\footnotesize{}
\begin{equation}
\mathbb{Q}=\left[\begin{matrix}-(q_{0,1}+q_{0,b}) & q_{0,1} & 0 & \cdots & \cdots & q_{0,b}\\
\\
\vdots &  & \ddots\\
 & 0 & q_{i,i-1} & -(q_{i,i-1}+q_{i,i+1}) & q_{i,i+1}\\
\vdots &  &  &  & \ddots
\end{matrix}\right],\label{reactmat2}
\end{equation}
}where the transition rates are given by 
\begin{equation}
q_{i,i\pm1}=\frac{D}{\delta r^{2}}\pm\frac{D}{r_{i\pm1}\delta r},\label{probdiff2}
\end{equation}
and $q_{0,b}=\tilde{\kappa}(r)$. Note the rows of $\mathbb{Q}$ now
sum to zero as we should expect from a continuous time Markov chain.
The $i^{th}$ equation has the form, 
\begin{equation}
\frac{d\pi_{i}(t)}{dt}=q_{i-1,i}\pi_{i-1}(t)-(q_{i,i-1}+q_{i,i+1})\pi_{i}(t)+q_{i+1,i}\pi_{i+1}(t).\label{eq:SME-ind}
\end{equation}

Note we are truncating the system up to shell $N$ (the $(N+1)^{th}$
column of matrix $\mathbb{\mathbb{Q}}$) and that other discretizations
are possible \citep{wang2003robust}. To strictly recover Eq. (\ref{eq:smol2BC}),
we need $N\rightarrow\infty$. However it is simpler to add $q_{0,b}$
at the end of the first row of the matrix, which corresponds to a
periodic boundary condition. This means every time a particle reacts
at $\sigma$, a new one is placed at the outermost shell $r_{N}$.
This periodic condition will be consistent with the model from Appendix
\ref{sec:canon}; however, it will not be necessary in Sec. \ref{sec:grand}. 

This model describes the probability distribution dynamics of one
$B$ molecule diffusing around an $A$ molecule with a reaction boundary
at $\sigma$; it is the discrete analog of the probabilistic approach
from Sec. \ref{sec:difreat-prob}, and it recovers Eq \ref{eq:smol2}
in the continuous limit \citep{del2016discrete}. In the next section,
we will construct another discrete-state probabilistic model that
is the analog of the concentration-based approach of Sec. \ref{sec:difreac}.
The advantage of employing a discrete state space will become evident
in Secs. \ref{sec:grand}--\ref{sec:sim_exocyt}.

\section{Grand canonical Smoluchowski master equation \label{sec:grand} }

In order to provide a complete probabilistic interpretation of Smoluchowski's
original concentration-based approach from Sec. \ref{sec:difreac},
we need to first generalize the SME to an arbitrary number $m$ of
$B$ particles in the system. A simple generalization is achieved
by assuming the total number of particles $m$ remains constant (canonical
ensemble), see Appendix \ref{sec:canon}. In this section, we present
the GC-SME, a generalization of the SME to a nonconstant total number
of particles $m$ (grand canonical ensemble). In Sec. \ref{subsec:hydlim}
and Appendix \ref{sec:largecopylim}, we show the GC-SME recovers
the concentration-based approach through two different limiting behaviors,
resembling Kurtz limit \citep{kurtz1971limit,kurtz1972relationship},
where the mass-action ODE is recovered as a limiting case of the chemical
master equation. These results will establish the
connection between the probabilistic approach of Sec. \ref{sec:difreat-prob}
and the concentration-based approach of Sec. \ref{sec:difreac}. A
diagram summarizing the connections between the different models is
shown in Fig. \ref{fig:theorydiag}. In Sec. \ref{subsec:SMinterpret},
we provide a statistical mechanical interpretation of the concentration-based
model based on the GC-SME in order to clarify interpretations at the
particle level. Sec. \ref{subsec:chempot} uses the GC-SME to bridge
the concept of chemical potential from a probabilistic level to a
framework of densities.

We begin with the construction of the GC-SME. Analogous
to Sec. \ref{sec:SME}, we consider a macromolecule $A$ fixed at
the origin and partition the space around $A$ in spherical shells
of width $\delta_{r}$. The dynamics of the $B$ particles are described
by a master equation for the joint probability of having $n_{i}$
$B$ particles in shell $i$, $P(n_{0},n_{1},\dotsc n_{N},t)$, where
$\sum_{i=0}^{N}n_{i}=m$ is not constant over time. Each particle
will diffuse following the SME dynamics, i.e. the rates from
Eq. (\ref{eq:SME}) and (\ref{probdiff2}). The inner boundary is a partially
absorbing boundary; however it could be replaced by other reaction
processes, like a purely absorbing boundary or a volume reactivity
model \citep{doi1976stochastic}. The outer boundary allows particles
to diffuse out of shell $i=N$ and be annihilated. We further introduce
a material bath by adding an additional outer shell $i=N+1$ with
a constant number of particles that can diffuse into shell $i=N$.
With these considerations, we can write the GC-SME
{\small
\begin{align}
 & \frac{d}{dt}P({\scriptstyle n_{0},n_{1},\dotsc,n_{N},t})=\nonumber \\
{\scriptstyle \text{reaction boundary}} & \begin{cases}
P({\scriptstyle n_{0}-1,n_{1},\dotsc,n_{N},t})q_{-1,0}({\textstyle n_{-1})}\\
+P({\scriptstyle n_{0}+1,n_{1},\dotsc,n_{N},t})q_{0,-1}(n_{0}+1)
\end{cases}\nonumber \\
{\scriptstyle \text{inner diffusion}} & \begin{cases}
+P({\scriptstyle n_{0}+1,n_{1}-1,\dotsc,n_{N}})q_{0,1}(n_{0}+1)\\
+P({\scriptstyle n_{0}-1,n_{1}+1,\dotsc,n_{N}})q_{1,0}(n_{1}+1)\\
+P({\scriptstyle n_{0},n_{1}+1,n_{2}-1\dotsc,n_{N}})q_{1,2}(n_{1}+1)\\
+P({\scriptstyle n_{0},n_{1}-1,n_{2}+1\dotsc,n_{N}})q_{2,1}(n_{2}+1)\\
\vdots\\
+P({\scriptstyle n_{0},\dotsc,n_{N-1}+1,n_{N}-1})q_{N-1,N}(n_{N-1}+1)\\
+P({\scriptstyle n_{0},\dotsc,n_{N-1}-1,n_{N}+1})q_{N,N-1}(n_{N}+1)
\end{cases}\nonumber \\
{\scriptstyle \text{outer boundary}} & \begin{cases}
+P({\scriptstyle n_{0},\dotsc,n_{N}+1})q_{N,N+1}(n_{N}+1)\\
+P({\scriptstyle n_{0},\dotsc,n_{N}-1})q_{N+1,N}(n_{N+1})
\end{cases}\nonumber \\
{\scriptstyle \text{leaving state}} & \begin{cases}
{\displaystyle -P({\scriptstyle n_{0},\dotsc,n_{N}})\sum_{k=0}^{N}\left[q_{k,k+1}+q_{k,k-1}\right]n_{k}.}\end{cases}\label{eq:GCE-mastereq}
\end{align}
}
We divided the terms of the GC-SME into four categories: the incoming
transitions to the current state through the reaction boundary, the
incoming transitions to the current state through diffusion of particles
in the inner shells, the incoming transitions to the current state
through the outer boundary in contact with a material bath and the
transitions leaving the current state through diffusion or escape
through either of the boundaries.

\subsection{Hydrodynamic limit \label{subsec:hydlim}}

In this section, we show the mean-field of the GC-SME
recovers the concentration-based approach in the continuous limit
(hydrodynamic limit), finalizing a clear-cut connection between the
probabilistic and the concentration-based approach (Fig. \ref{fig:theorydiag}).

We obtain the mean-field of the GC-SME by deriving
an equation for the expected number of molecules at shell $i$. Multiplying
the GC-SME by $n_{i}$, summing over all the possible number of molecules
$\{\bar{n}\} = \{n_0,\dotsc,n_N\}$ for all $j=0,1,2\dotsc$, using that $\left\langle n_{i}\right\rangle =\mathbb{E}[\mathcal{N}_{i}=n_{i}]$
and doing some algebra, we obtain the mean-field equation
\begin{align*}
\sum_{\{\bar{n}\}} & n_{i}\frac{d}{dt}P(n_{0},n_{1},\dotsc,n_{N},t)=\frac{d\left\langle n_{i}\right\rangle }{dt}=\\
 & \left\langle n_{i+1}\right\rangle q_{i+1,i}-\left\langle n_{i}\right\rangle \left[q_{i,i+1}+q_{i,i-1}\right]+\left\langle n_{i-1}\right\rangle q_{i-1,i}\\
 & \underbrace{\qquad+\sum_{j=-1}^{N}\left[\left\langle n_{i}n_{j}\right\rangle q_{j,j+1}+\left\langle n_{i}n_{j+1}\right\rangle q_{j+1,j}\right]\qquad}_{\text{absorbing boundary }+\text{ inner diffusion }+\text{ outer boundary}}\\
 & \underbrace{-\sum_{j=0}^{N}\left\langle n_{i}n_{j}\right\rangle \left[q_{j,j+1}+q_{j,j-1}\right]-\left\langle n_{i}n_{N+1}\right\rangle q_{N+1,N}}_{\text{leaving state}}.
\end{align*}
As $q_{-1,k}=0$ for all $k$, and $q_{0,-1}=q_{0,b}$, we can join
the two series together. All the terms in the series will cancel out
except for one. This remaining term will also cancel out with the
second term from the ``leaving state''. The only terms left over
are 
\begin{equation}
\frac{d\left\langle n_{i}\right\rangle }{dt}=\left\langle n_{i+1}\right\rangle q_{i+1,i}-\left\langle n_{i}\right\rangle \left[q_{i,i+1}+q_{i,i-1}\right]+\left\langle n_{i-1}\right\rangle q_{i-1,i}.\label{eq:FD-expval2}
\end{equation}
Renaming $F_{i}(t)=\left\langle n_{i}\right\rangle $, we have exactly
the same equation as Eq. (\ref{eq:FD-expval}), so we will have the
same limiting behavior as the one studied in detail in Appendix \ref{sec:canon}.
We will now take the continuous limit. We first substitute
the transition rates from Eq. (\ref{probdiff2}) to obtain
{\small{}
\begin{align}
\frac{dF_{i}(t)}{dt}=D\left[\frac{F_{i+1}(t)-2F_{i}(t)+F_{i-1}(t)}{\delta r^{2}}\right]\nonumber \\
-\frac{2D}{r_{i}}\left[\frac{F_{i+1}(t)-F_{i-1}(t)}{2\delta r}\right]+\frac{D}{\delta r}\left[\frac{F_{i}(t)}{r_{i}-\delta r}-\frac{F_{i}(t)}{r_{i}+\delta r}\right],\label{eq:ithSME-1}
\end{align}
which is the same as Eq. (\ref{eq:ithSME}). Taking
the limit $\delta r\to0$ and scaling the geometrical effects (Eq.
(\ref{eq:spscaling})), we recover the Smoluchowski equation (Eq.
\ref{eq:smol}) 
\[
\frac{\partial f(r,t)}{\partial t}=\frac{D}{r^{2}}\frac{\partial}{\partial r}\left(r^{2}\frac{\partial f(r,t)}{\partial r}\right),
\]
where the function $f(r,t)$ is the expected value for the concentration.
However, in this case, the interesting behavior will be at the boundaries
($i=0,\,N$ in Eq. (\ref{eq:FD-expval2})). The resulting equations
are 
\begin{align}
\frac{dF_{0}(t)}{dt} & =-(q_{0,1}+q_{0,b})F_{0}(t)+q_{1,0}F_{1}(t),\label{eq:GCSMEinnerBC}\\
\frac{dF_{N}(t)}{dt} & =q_{N+1,N}n_{N+1}-(q_{N,N+1}+q_{N,N-1})F_{N}(t)\label{eq:eq:GCSMEouterBC}\\
 & +q_{N-1,N}F_{N-1}(t),\nonumber 
\end{align}
where we used the fact that the number of particles at $i=N+1$ is
fixed , $\left\langle n_{N+1}\right\rangle =n_{N+1}$. In
the continuous limit, in the same way as Eq. (\ref{eq:FD-innerBC}),
the equation for the inner boundary, Eq. (\ref{eq:GCSMEinnerBC}),
will yield the boundary condition for the inner absorbing boundary
\[
\left.4\pi D\sigma^{2}\frac{\partial f(r,t)}{\partial r}\right|_{r=\sigma}=\kappa f(\sigma,t).
\]
In order to determine the boundary condition in the far-field, we
apply the same methodology to Eq. (\ref{eq:eq:GCSMEouterBC}). We
introduce a ghost cell at $i=N+1$, $\tilde{F}_{N+1}$. By
adding and subtracting terms with $\tilde{F}_{N+1}$, we rewrite Eq.
(\ref{eq:eq:GCSMEouterBC}) in such a way that we have all the terms
from Eq. (\ref{eq:ithSME-1}) plus some additional terms. Writing
the rates explicitly, we obtain 
\begin{align*}
\frac{dF_{N}(t)}{dt}=D\left[\frac{\tilde{F}_{N+1}(t)-2F_{N}(t)+F_{N-1}(t)}{\delta r^{2}}\right]\\
+\frac{D}{\delta r}\left[\frac{F_{N}(t)}{r_{N}-\delta r}-\frac{F_{N}(t)}{r_{N}+\delta r}\right]-\frac{D}{\delta r^{2}}\tilde{F}_{N+1}(t)+n_{N+1}q_{N+1,N}\\
+\tilde{F}_{N+1}(t)\frac{D}{r_{N}\delta r}-\frac{2D}{r_{N}}\left[\frac{\tilde{F}_{N+1}(t)-F_{N-1}(t)}{2\delta r}\right]
\end{align*}
In order to satisfy the main equation (Eq. (\ref{eq:ithSME-1})),
the additional terms must be zero, so the ghost cell needs to satisfy
\[
n_{N+1}q_{N+1,N}=\left[\frac{D}{\delta r^{2}}-\frac{D}{r_{N}\delta r}\right]\tilde{F}_{N+1}(t).
\]
We apply directly the scaling from Eq. (\ref{eq:spscaling}); however,
in this case $F_{i}(t)$ is still discrete, so $F_{i}(t)=4\pi r_{i}^{2}f(r_{i},t)\delta r$.
Additionally, the concentration $c_{0}$ in the outermost shell is
given by $c_{0}=n_{N+1}/4\pi r_{N+1}^{2}\delta r$. We will also call
the rate of incoming particles $\gamma=q_{N+1,N}$. Substituting these
into the equation, we obtain
\begin{align*}
c_{0}4\pi r_{N+1}^{2}\delta r\gamma= & \left[\frac{D}{\delta r^{2}}-\frac{D}{r_{N}\delta r}\right]4\pi r_{N+1}^{2}\tilde{f}(r_{N+1},t)\delta r,\\
\Rightarrow\ \ \ \ \ c_{0}\delta r^{2}\gamma= & \left[1-\frac{\delta r}{r_{N}}\right]D\tilde{f}(r_{N+1},t).
\end{align*}
In order to obtain a convergent limit, we need to set the transition
rate $\gamma$ to have the value 
\begin{equation}
\gamma=\frac{D}{\delta r^{2}}-\frac{D}{r_{N}\delta r},\label{eq:gammmRate}
\end{equation}
which yields $c_{0}={\tilde{f}(r_{N+1},t)}$.
The limit as $\delta r\to0$ yields 
\[
f(r_{\text{max}},t)=c_{0},
\]
where the number of particles $n_{N+1}$ corresponds to a constant
bath concentration of $c_{0}$. The rate $\gamma$
corresponds to the diffusion rate for the corresponding discretization,
which is intuitively consistent. Other values of $\gamma$ could be
provided in the discrete model, but they will not produce the correct
continuous limit. This result shows the concentration-based approach,
with its corresponding boundary conditions, is recovered in the hydrodynamic
limit of the GC-SME; it is referred to as hydrodynamic limit following
the literature of interacting particle systems \citep{franco2014interacting,kipnis2013scaling}.

Note there was a hidden assumption when we stated
$c_{0}=n_{N+1}/4\pi r_{N+1}^{2}\delta r$. The state $N+1$
is different to all the others since the number of particles does
not change even though the system is continually absorbing particles
from it. This is only feasible if it can access an infinite number
of particles. When assuming a constant concentration $c_{0}=n_{N+1}/4\pi r_{N+1}^{2}\delta r$,
we actually refer to the concentration of the whole bath 
\[
c_{0}=\frac{3n_{\text{bath}}}{4\pi(R_{\infty}^{3}-r_{N}^{3})}.
\]
In order to have access to an infinite amount of particles, we need
to make the corresponding volume infinite, $R_{\infty}\to\infty$.
At the boundary layer of width $\delta r$ around $r_{N}$, the concentration
has to be $c_{0}=n_{N+1}/4\pi r_{N+1}^{2}\delta r$, where $n_{N+1}\to0$
as $\delta r\to0$. Although it might appear the number of particles
of the bath goes to zero, it is actually the opposite; the number
of particles and the volume in the bath must go to infinity at a fixed
rate. 

In addition to the hydrodynamic limit, one can also
show that the concentration-based approach is also recovered in the
large copy number limit of the GC-SME; this result is shown in Appendix
\ref{sec:largecopylim}. 

With these results, we finalize the connection between
the probabilistic and the concentration-based approach, see Fig. \ref{fig:theorydiag}.
Smoluchowski's original concentration-based approach is therefore
better understood in terms of a probabilistic model in two different
ways, as the hydrodynamic limit or as the large copy number limit
of the GC-SME. Note the latter does not require taking the mean field.
However, it is not surprising these two limits converge to the same
result since this is a linear system.

These results will further allow us to generate
particle-based simulations that are consistent with concentration
descriptions (Secs. \ref{sec:app_nummeth} and \ref{sec:sim_exocyt}).
The results of one of the simulations is shown in Fig. \ref{fig:LLN_conv},
where the two types of convergence of the GC-SME are illustrated with
particle-based simulations.

We should also emphasize the relevance of moving
into a discrete state setting since it is not clear how one could
write the GC-SME using a continuous state spectrum.

\begin{figure}
\includegraphics[scale=0.7]{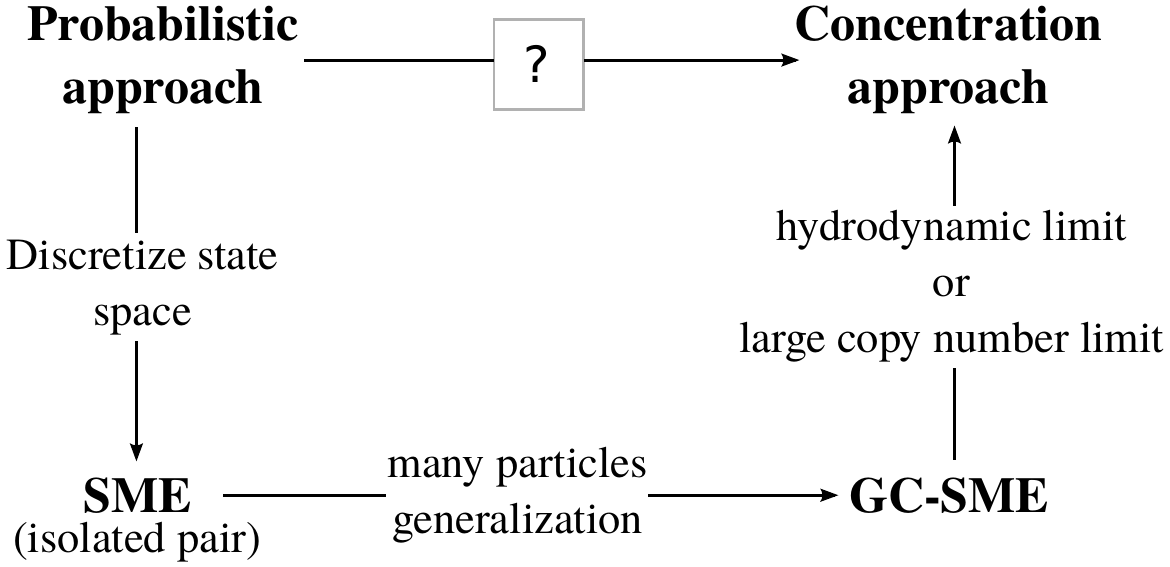}

\caption{Diagram showing the relationships between the different
models. The initial unknown is how Smoluchowski's probabilistic and
concentration-based approaches relate to each other. In order to resolve
this, we first derive the SME, which is a state-space discretization
of the probabilistic approach. Then we derive the GC-SME, a stochastic
model that generalizes the SME to a large number of B particles (non-constant).
Finally, the concentration-based model is recovered as the hydrodynamic
limit (mean-field) or large copy number limit of the GC-SME. As in
particle-based simulations, one requires to know the dynamics of the
stochastic trajectories, this connection allows developing particle-based
simulations that are consistent with bulk concentration dynamics.}

\label{fig:theorydiag}
\end{figure}

\begin{figure*}[t]
\centering \includegraphics[width=0.32\textwidth]{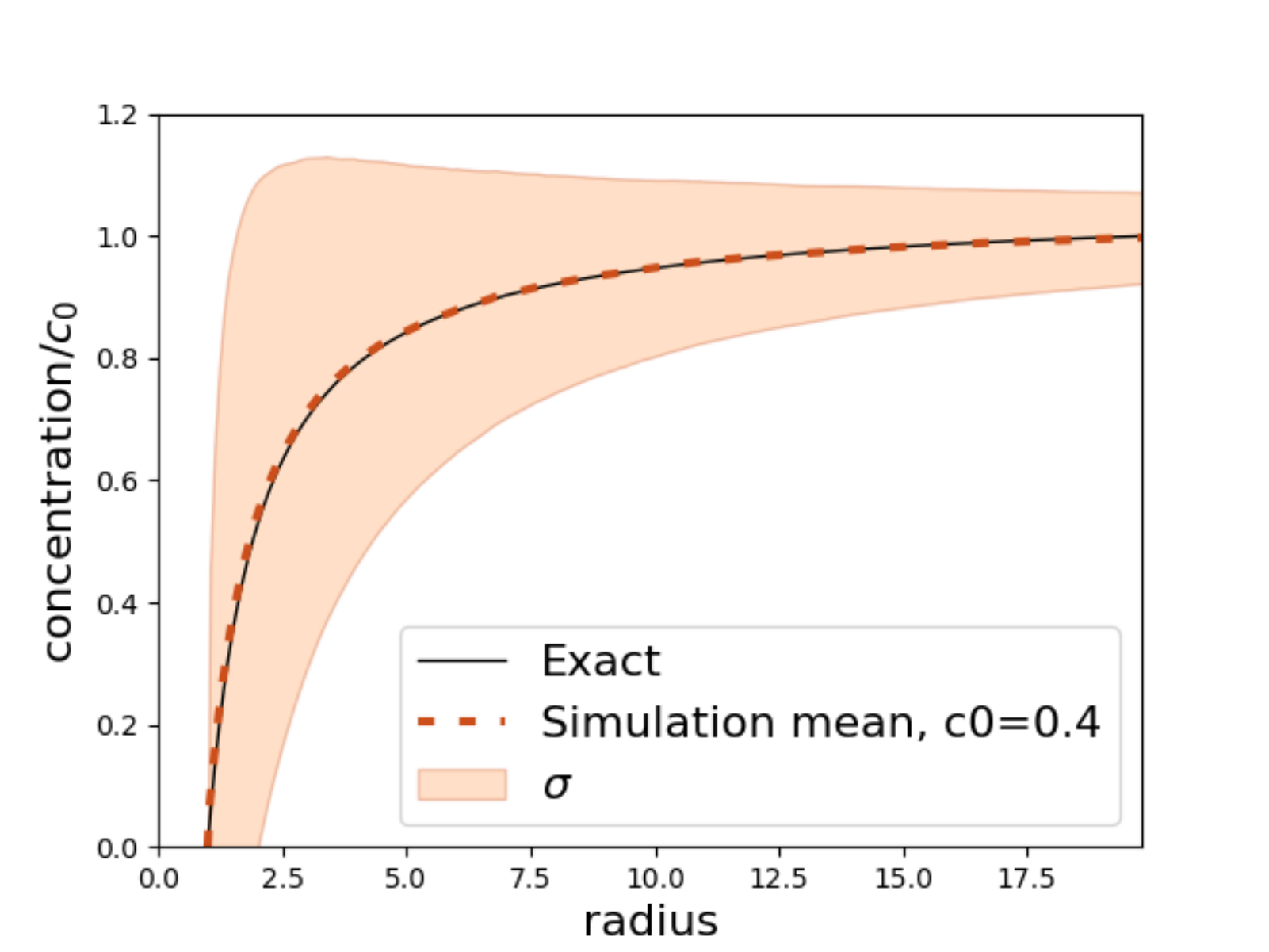}
\includegraphics[width=0.32\textwidth]{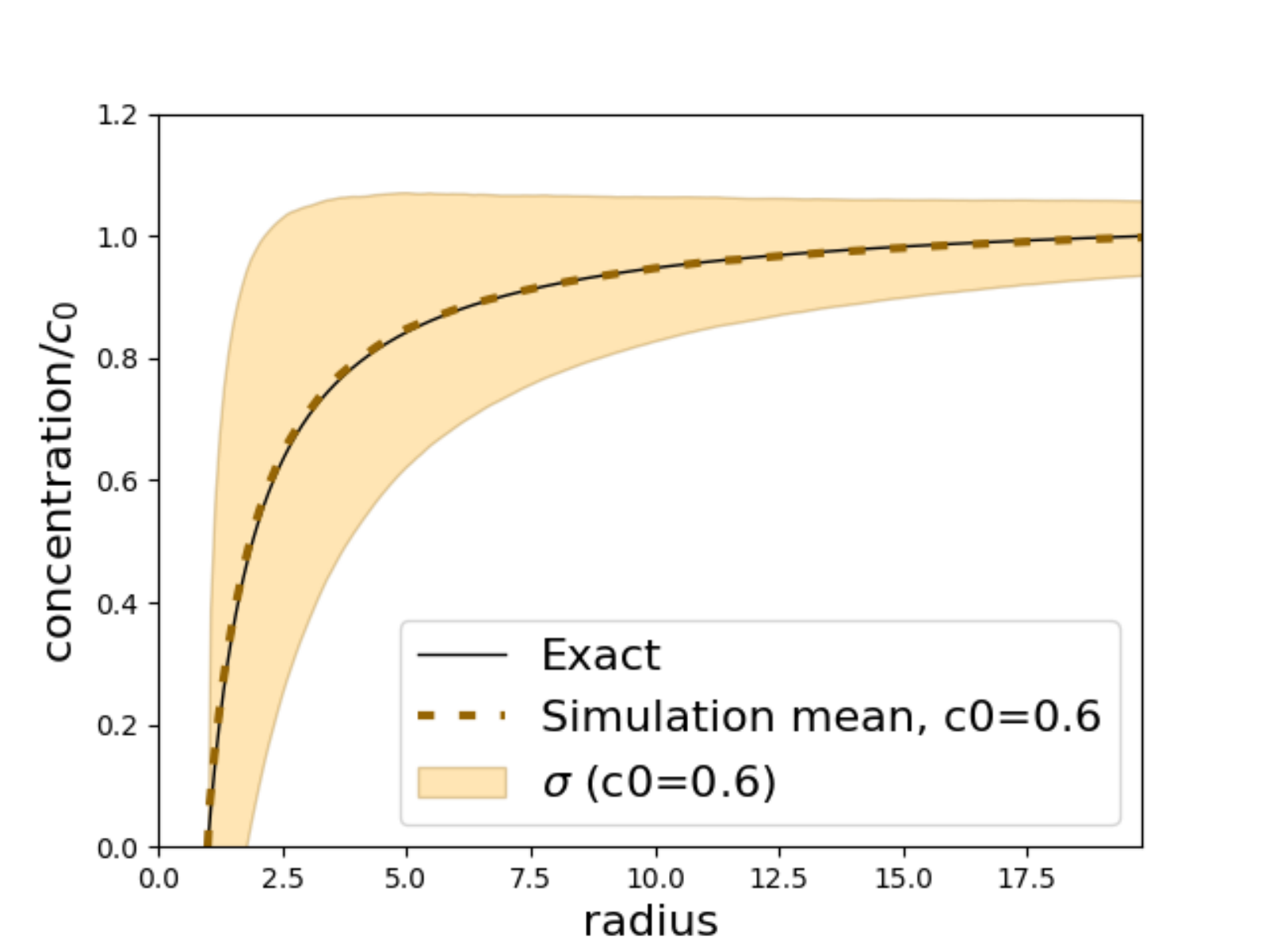}
\includegraphics[width=0.32\textwidth]{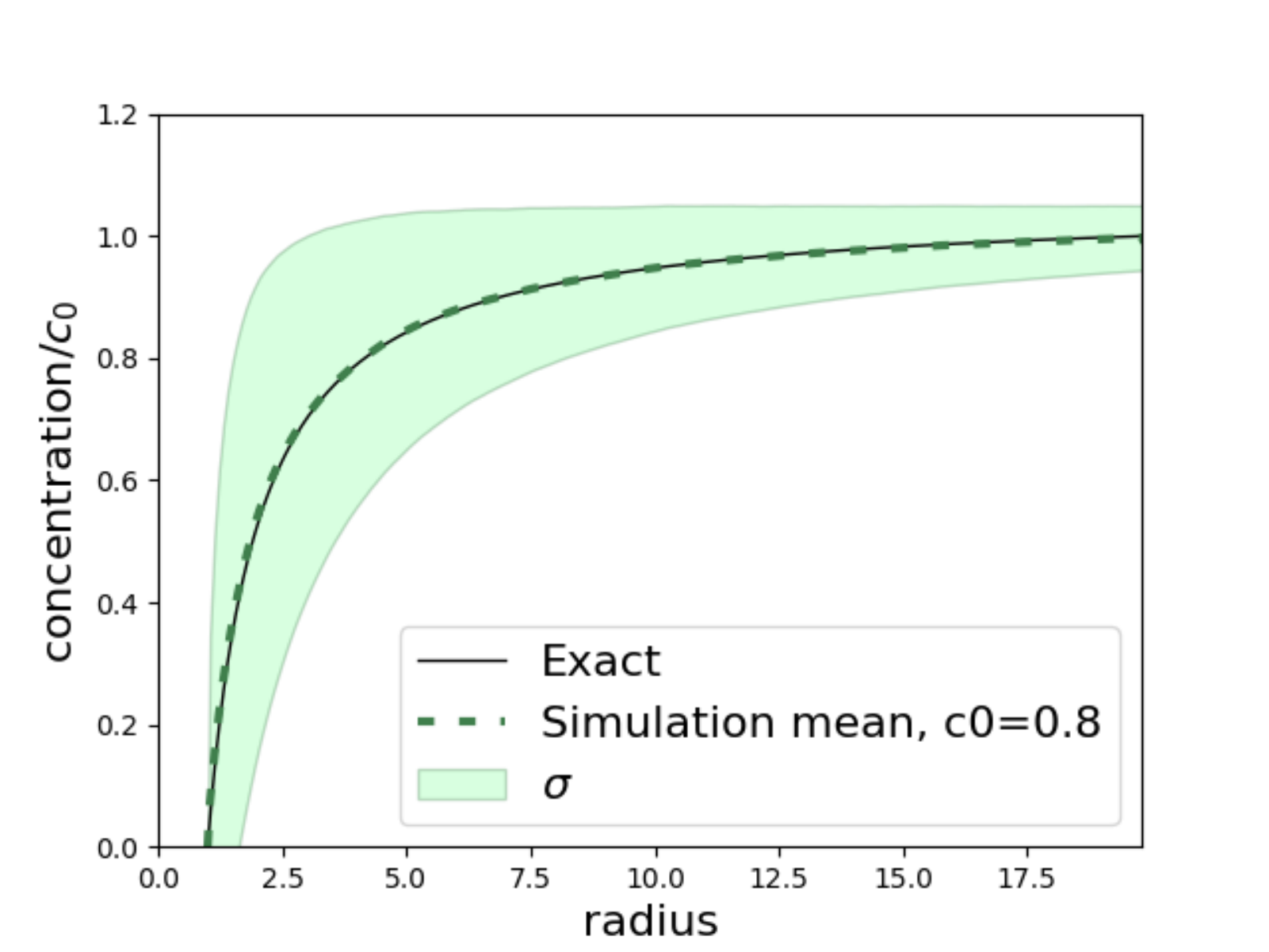} 

\includegraphics[width=0.32\textwidth]{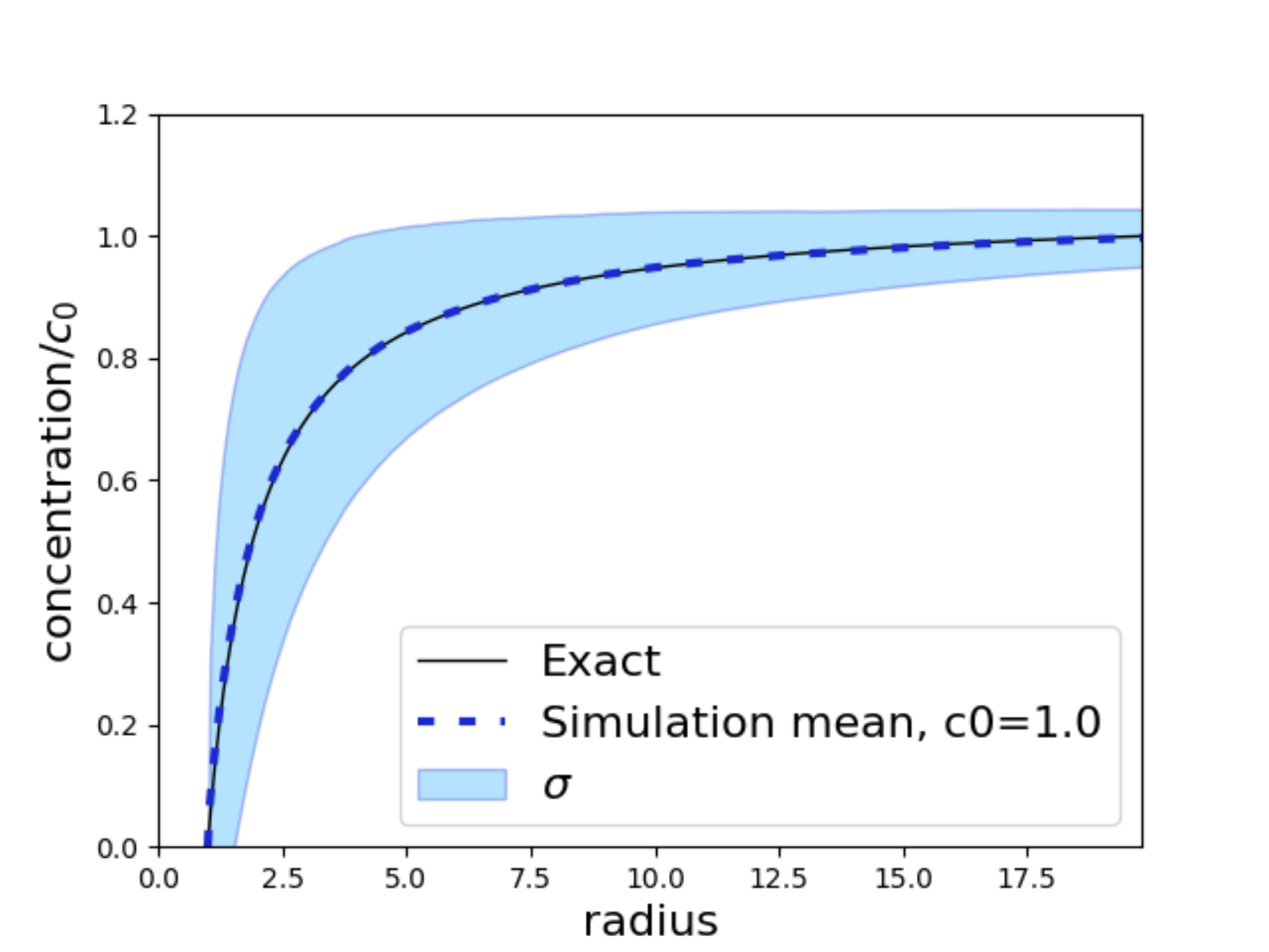}
\includegraphics[width=0.32\textwidth]{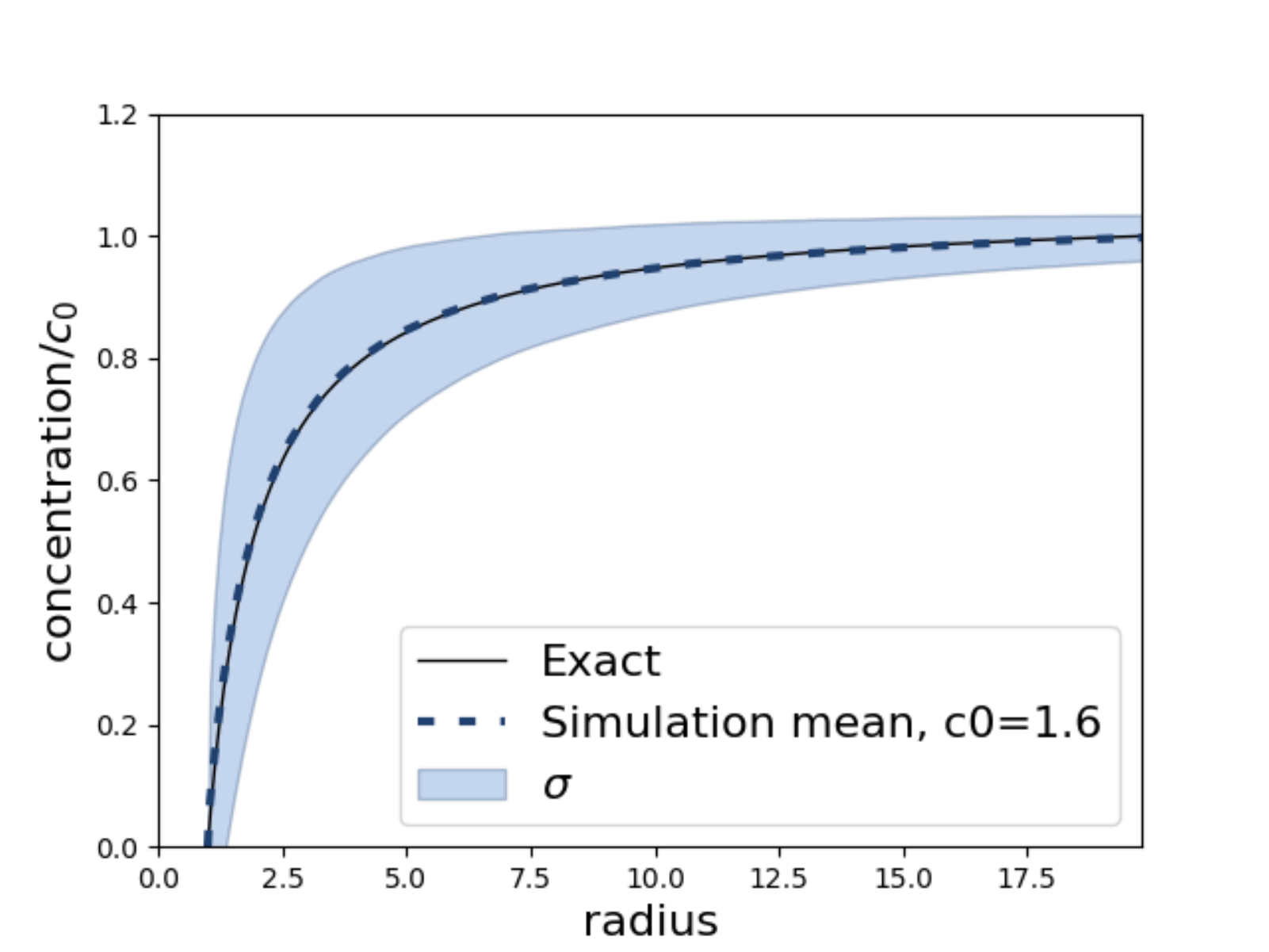}
\includegraphics[width=0.32\textwidth]{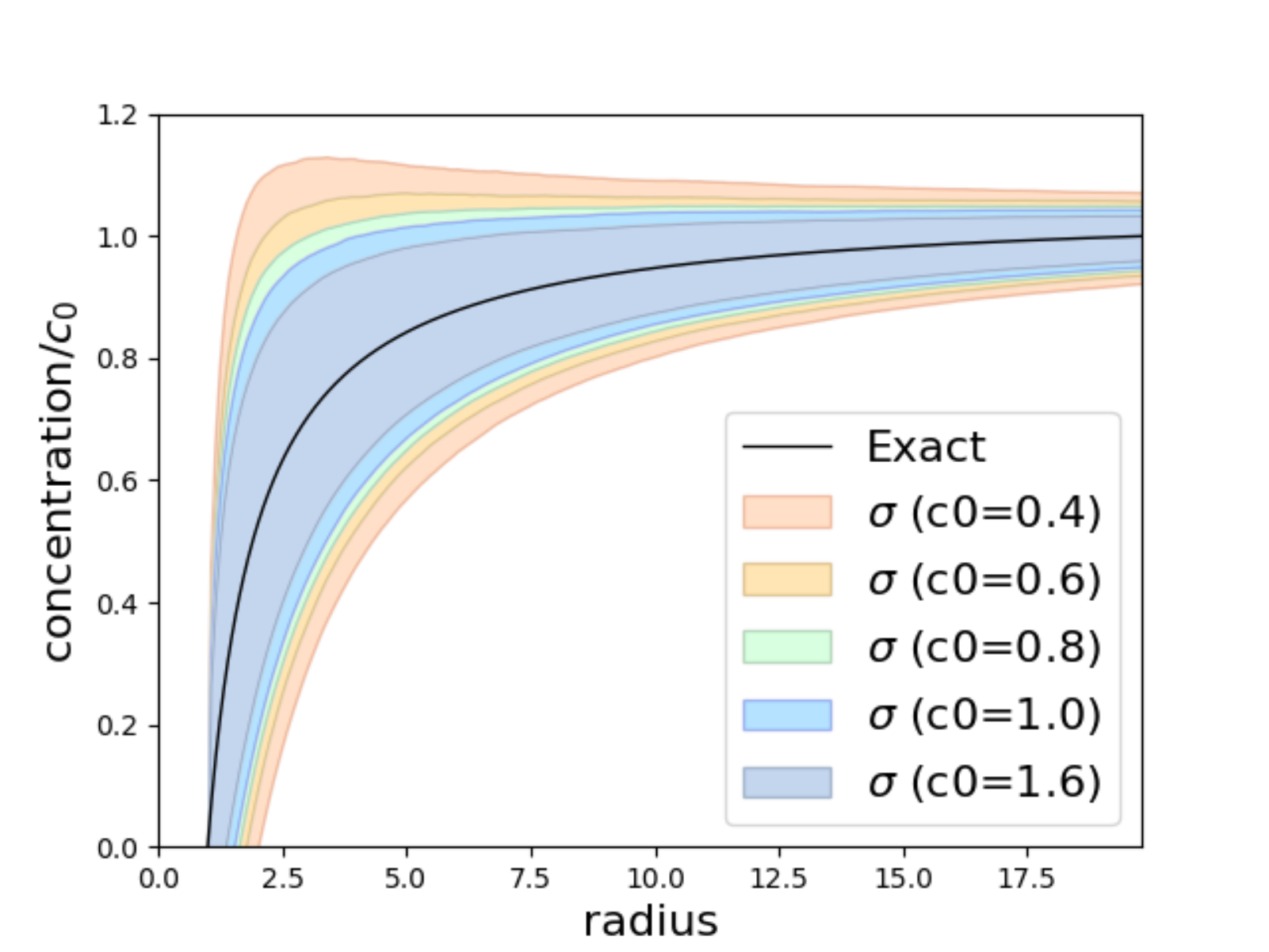}
\caption{Comparison between the exact solution of Smoluchowski's original concentration-based
approach and particle-based simulations based on the GC-SME (purely-absorbing
boundary). Results are plotted for five different values of the bath
concentration $c_{0}=[0.4,0.6,0.8,1.0,1.6]$; the standard deviation is represented by the shaded regions. The simulations were
performed following the methodology described in Sec. \ref{sec:app_nummeth},
and they were averaged over $6\times10^{6}$ time steps, with $D=5$,
$dt=0.0002$, $\sigma=1$ and $r_{\text{max}}=20$.
The results were normalized dividing by $c_{0}$, so these graphs
are representative of $\mathcal{N}_{j}^{\text{fr}}$ (see Appendix \ref{sec:largecopylim}). The last plot
shows the standard deviations for the five concentrations on top of
each other for comparison. This is a visual representation of the
two different types of convergence: the mean field convergence from
the hydrodynamic limit, where the mean matches the exact solution
regardless of the bath state concentration; and the large copy number
limit convergence (Eq. (\ref{eq:GC_lln})), where the standard deviation
is consistently reduced as the total number of particles increases
($c_{0}\rightarrow\infty$).}
\label{fig:LLN_conv} 
\end{figure*}

\subsection{Statistical mechanical interpretation \label{subsec:SMinterpret} }

We showed that Smoluchowski's original concentration-based approach
emerges from the hydrodynamic/large copy number limit of the GC-SME.
This connection yields specific interpretations of Smoluchowski's
original model. In order to elucidate them, first note there are two
equivalent interpretations of a system in the grand canonical ensemble
\citep{pathria1996statistical}:
\begin{enumerate}
\item The system is immersed in a large reservoir with which it can exchange
energy and particles. 
\item The given system and a large number of \textquotedblleft hypothetical
copies'' can exchange energy and particles with each other. 
\end{enumerate}
Smoluchowski's concentration-based approach is framed
following 1; there is one macromolecule $A$ in the origin surrounded
by a concentration gradient of $B$ ligands. $A$ acts as a sink of
ligands since it can react with an infinite number of them. In most
realistic settings, a macromolecule can only react with one or a finite
number of ligands, so why Smoluchowski's original approach models
so many systems successfully? If we concentrate on the GC-SME instead
of the original concentration approach, given that $B$ particles
are treated as independent entities, we can easily frame the model
following either interpretation 1 or 2. The GC-SME therefore allows
us to interpret Smoluchowski's concentration-based approach following
2: consider an ensemble of systems each with one macromolecule $A$
and all inside some solution. Each system in our ensemble corresponds
to a small neighborhood around each $A$, where there are no other
$A$'s. The $B$ ligands in the solution are plentiful and can diffuse
through the whole solution. Therefore, they can be exchanged between
the different systems of our ensemble. This description does not require
one $A$ to react with a large number of $B$'s because our ensemble
has a large number of $A$'s as well. This means
that the concentration gradient resulting from Smoluchowski's original
theory is the average concentration we would observe when looking
around each one of the $A$'s in the solution. 

Although this interpretation was previously stated
in \citep{agmon1990theory,schurr-1970}, the GC-SME further provides
a precise probabilistic interpretation, enabling explicit implementation
of particle exchange mechanisms that are consistent with concentration-based
models. It should be pointed out that different exchange mechanisms
at the particle level could potentially yield the same mean-field
behavior; however, physical arguments on a case by case basis can
be used to discard alternative mechanisms. In our case, the particles
are injected into the system following a Poisson process with a constant
rate along with a first-order exit rate; this is physically consistent
with the modeling of diffusion with a Markov model.

\subsection{Emergent macroscopic chemical potential \label{subsec:chempot}}

The chemical potential summarizes the thermodynamics arising from
diffusion and reaction in a chemical system. Deriving the chemical
potential for diffusion-influenced reactions could help reconcile
nonequilibrium thermodynamics with chemical reactions \citep{bedeaux2010mesoscopic}.

In order to derive the chemical potential, we begin by calculating
a generalized version of Gibbs equilibrium free energy
\citep{ge2016mesoscopic,ge2017mathematical}

\[
\phi^{ss}=-k_{B}T\ln{P^{ss}},
\]
where $P^{ss}$ is the steady-state probability distribution, $k_{B}$
is the Boltzmann constant and $T$ the temperature. We will first
calculate this function in terms of the number of particles, and then
we will proceed to a concentration description. We
substitute the (nonequilibrium) steady state distribution $P^{ss}(n_{0},n_{1},\cdots,n_{N})$,
which in the Appendix \ref{sec:largecopylim} is shown to satisfy
Eq. (\ref{eq:p_poisson}). As this is only a product of marginal
distributions, we can expand it as 
\begin{align}
\phi^{ss}(n_{0},n_{1},\cdots,n_{N})=-k_{B}T\sum_{i=0}^{N}\ln\left(P^{ss}(n_{i})\right),\label{eq:phiss_01}
\end{align}
where $P^{ss}(n_{i})=\left\langle n_{i}\right\rangle e^{-\left\langle n_{i}\right\rangle }/n_{i}!$
is the marginal steady state probability of having $n_{i}$ particles
on shell $i$ at time $t$. We can do a continuous interpolation by
using the gamma function instead of the factorial (note the integral
of the continuous distribution integrates to one), 
\begin{align*}
P^{ss}(n_{i})=\frac{\left\langle n_{i}\right\rangle ^{n_{i}}}{\Gamma[n_{i}+1]}e^{-\left\langle n_{i}\right\rangle }\approx\frac{1}{\sqrt{2\pi}}\left[\frac{\left\langle n_{i}\right\rangle }{n_{i}}\right]^{n_{i}}e^{n_{i}-\left\langle n_{i}\right\rangle }
\end{align*}
where for the second equality we simply used Stirling's approximation
$\Gamma(k+1)\approx\sqrt{2\pi k}(k/e)^{k}$ and $n_{i}+1/2\approx n_{i}$,
which are both valid for $n_{i}\gg1$. Inserting this again into Eq.
(\ref{eq:phiss_01}), we obtain 
\begin{gather*}
\phi^{ss}({\scriptstyle n_{0},n_{1},\cdots,n_{N}})\approx -k_{B}T\sum_{i=0}^{N}\ln\left(\frac{1}{\sqrt{2\pi}}\left[\frac{\left\langle n_{i}\right\rangle }{n_{i}}\right]^{n_{i}}e^{n_{i}-\left\langle n_{i}\right\rangle }\right)\\
\approx k_{B}T\sum_{i=0}^{N}n_{i}\ln\left(\frac{n_{i}}{\left\langle n_{i}\right\rangle }\right)-n_{i}+\left\langle n_{i}\right\rangle +\ln\sqrt{2\pi}.
\end{gather*}
As this is an energy, the last term is just an irrelevant constant
factor. Furthermore, the number of particles $n_{i}$, can be simply
written in terms of the volume and concentration at shell $i$, $n_{i}=V_{i}c_{i}$,
so we can rewrite the equation as a function of the volumes and concentrations,
\begin{align*}
\phi^{ss}({\scriptstyle V_{0},c_{0},\cdots,V_{N},c_{N}})\approx k_{B}T\sum_{i=0}^{N}V_{i}\left[c_{i}\ln\left(\frac{c_{i}}{\left\langle c_{i}\right\rangle }\right)+\left\langle c_{i}\right\rangle -c_{i}\right].
\end{align*}
As the chemical potential is nothing more than the derivative of
the free energy with respect the concentration, $\mu(c_{i})=\partial/\partial c_{i}(\phi(V_{1}, c_{1},\cdots,V_{N}, c_{N}))$,
we simply obtain that the chemical potential at shell $i$ is 
\[
\mu(c_{i})\approx k_{B}T\ln\left(\frac{c_{i}}{\left\langle c_{i}\right\rangle }\right)^{V_{i}},
\]
where $\left\langle c_{i}\right\rangle =\int_{r_{i}}^{r_{i}+\delta r}4\pi r_{i}^{2}c(r)dr$
and $c(r)$ is the solution to the original Smoluchowski equation
for the concentration, Eq. (\ref{eq:smol}). The quantity $(c_{i}/\left\langle c_{i}\right\rangle )^{V_{i}}$
plays the role of the thermodynamic activity in this model. Note the
chemical potential dependence on $V_{i}$ is necessary since the chemical
potential depends on the spatial partition chosen and of the volume
of each shell in the partition (the concept of defining thermodynamic
quantities on partitions follows from nonequilibrium thermodynamics
theory). If we include the boundaries, the far-field boundary satisfies
$c_{N+1}=\left\langle c_{N+1}\right\rangle $ and consequently $\mu(c_{N+1})$=0.
On the other hand, the reaction boundary only absorbs molecules, which
leads to $\left\langle c_{-1}\right\rangle =\infty$ and $\mu(c_{-1})=-\infty$.
The systems is clearly a nonequilibrium open system driven by a chemical
potential difference between the material bath and the absorbing boundary.

This result bridges the concept of chemical potential at a probabilistic
level to a framework of densities, as previously done in \citep{ge2016mesoscopic,ge2017mathematical}.
However, this work takes it a step further establishing this connection
in a spatially inhomogeneous system with a simple reaction process.

One should note that the most relevant quantity is not the chemical
potential \textit{per se} but the chemical potential difference. To
establish the difference appropriately, we need to use a consistent
partition across the different states. For instance, consider partitions
with either radial intervals of equal length or equal volume in every
shell. In the latter case, the volume becomes irrelevant for the chemical
potential difference. Both partitions would yield different chemical
potential differences, but they would both describe correct dynamics
in their corresponding coordinates. 

\section{Particle-based simulations based on the GC-SME \label{sec:app_nummeth}}

\begin{figure*}[t]
\centering \textbf{a.} \includegraphics[width=0.2\textwidth]{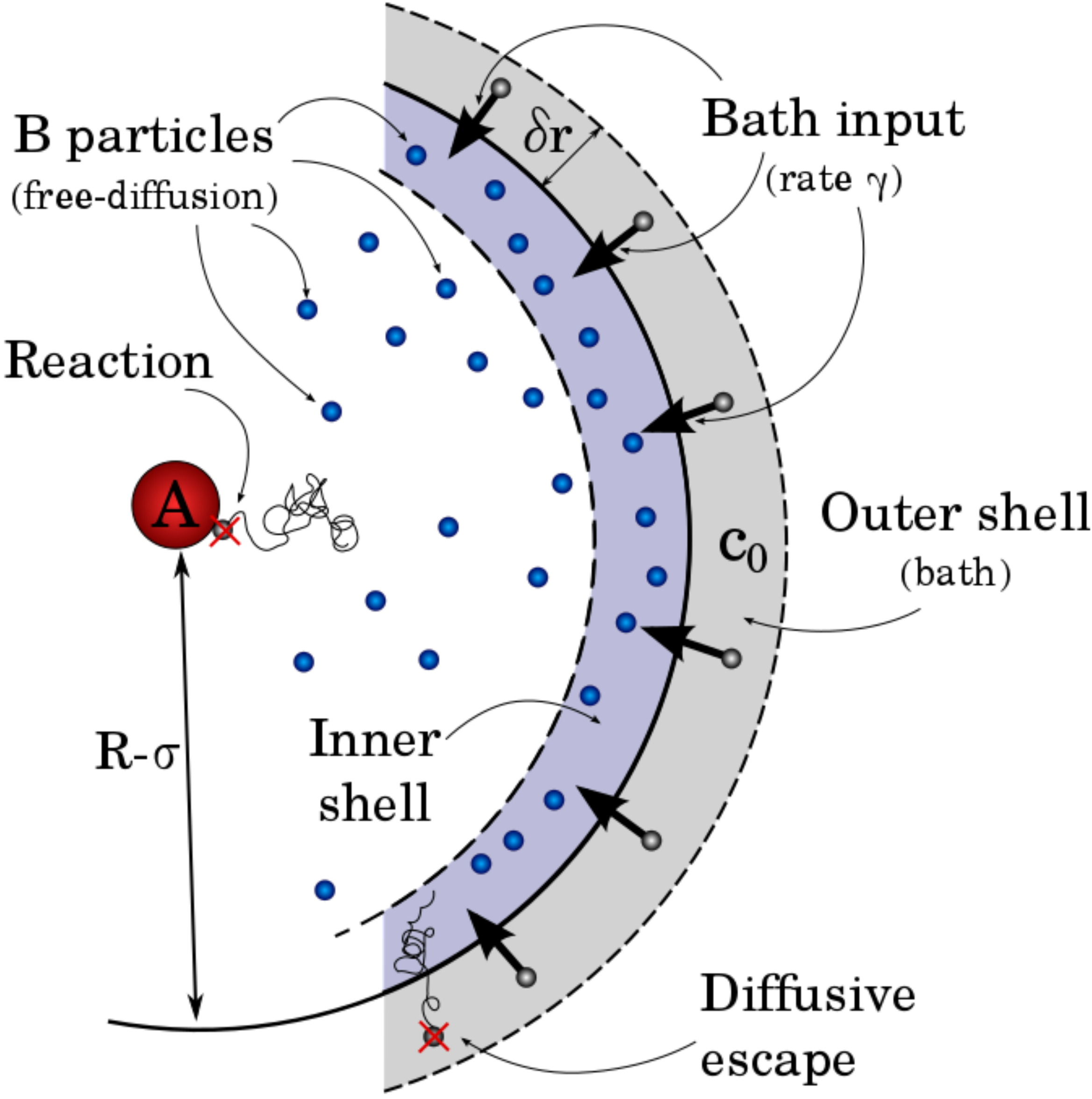}
\textbf{b.} \includegraphics[width=0.5\textwidth]{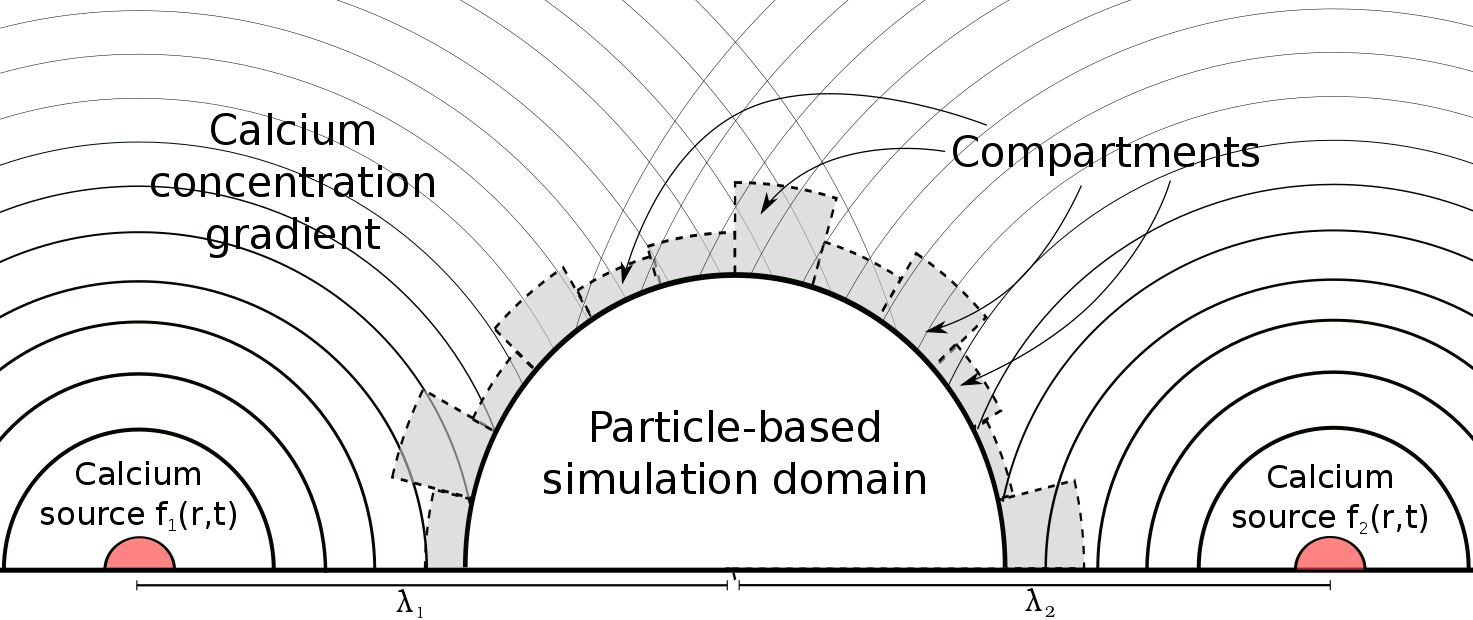}
\textbf{c.} \includegraphics[width=0.2\textwidth]{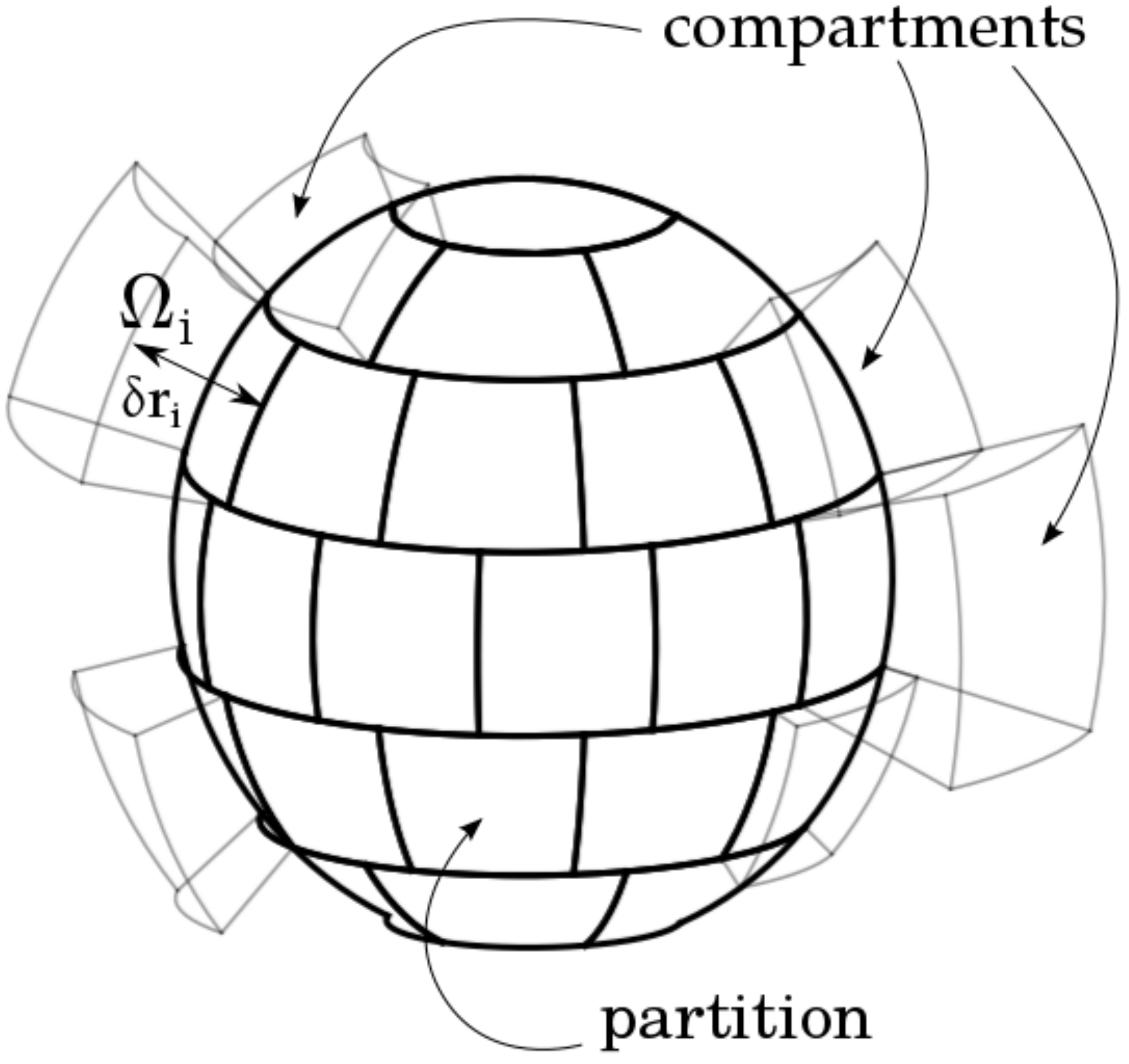}\\

\caption{Diagrams of particle-based simulations based on the GC-SME. \textbf{a.}
Diagram of a two-dimensional slice of the particle-based simulation
of Smoluchowski's model with constant concentration in the far-field
from Sec. \ref{sec:app_nummeth}. The gray-shaded shell $(R,R+\delta r]$
represents the material bath, where $n_{c_{0}}$ particles can jump
into the system at every time step. The blue-shaded shell $(R-\delta r,R]$
delimits where the particles from the bath can jump into. The $B$
particles (blue) can diffuse freely following standard Brownian motion
in $[\sigma,R]$. If a $B$ particle hits the reactive boundary at
$\sigma$ or escapes to $r>R$, it is eliminated. \textbf{b.} Diagram
of a two-dimensional slice of the model implemented in Sec. \ref{sec:sim_exocyt}.
Two sources placed at a distance $\lambda_{1}$ and $\lambda_{2}$
from the origin represent the calcium channels, which generate a concentration
gradient in the intracellular space. The particle-based simulation
is delimited by a half sphere surrounded by the compartments that
define the material bath. Each compartment has a number of particles
inside \texttt{$n_{i}$} that matches the corresponding average concentration
in the compartment at a given time. The compartments volumes vary to keep \texttt{$n_{i}$}
an integer. \textbf{c.} Three-dimensional illustration of the particle-based
simulation boundary showing an equal area partition of the half-sphere.
Each element of the partition has a corresponding $\delta r_{i}$,
which generates compartments with volume $\Omega_{i}$ to define the
material bath. For the sake of clarity, only a few compartments are
shown, and the compartment size variation is exaggerated.}
\label{fig:Gc_cartoon} 
\end{figure*}

In this section, we apply the previous results to produce arbitrary
particle-based simulations coupled to a constant concentration material
bath in the far-field. These simulations will be based on the GC-SME
from \ref{sec:grand}. Therefore, we consider a system where the mean
far-field ($r>R$) concentration $c_{0}$ of $B$ particles is constant.
We incorporate one $A$ particle at the origin with a purely absorbing
reactive boundary at $r=\sigma$ surrounded by spherical shells of
width $\delta r$ and an additional outer shell in the region $r\in[R,R+\delta r]$
to model the material bath. Analogous to the chemical master equation,
we can write the trajectory representation (Kurtz
representation \citep{anderson2015stochastic,kurtz1971limit,kurtz1972relationship})
of the GC-SME, which tracks the dynamics of the number of $B$ particles
$n_{i}(t)$ in shell $i$, 
\begin{align*}
n_{i}(t+\tau)=n_{i}(t)+\mathcal{R}_{i}(\{\bar{n}(t)\},\tau),
\end{align*}
where $\mathcal{R}(\{\bar{n}(t)\},\tau)$ denotes the random change
in the number of particles in shell $i$ and $\{\bar{n}(t)\}$ is the
set of elements $n_{i}(t)$ for every possible shell $i$. Naturally, it depends
on the time interval $\tau$ and the current state of the system $\{\bar{n}(t)\}$.
The process $\mathcal{R}_{i}(\{\bar{n}(t)\},\tau)$ is a composition
of two processes: 
\begin{itemize}
\item The diffusion of $B$ particles in the system, $\mathcal{D}_{i}(\{\bar{n}(t)\},\tau)$,
which includes diffusion across shells, reaction
by diffusion into the absorbing boundary ($r=\sigma$) and diffusion
out of the system into $r>R$. 
\item The injection of $B$ particles coming from the material bath (outer
shell) into the system, $\mathcal{I}_{i}(\{\bar{n}(t)\},\tau)$. 
\end{itemize}
These two processes are in general coupled and can occur at different
time-scales, so the time integration requires a robust scheme, like
Strang splitting \citep{kim2017stochastic}. The Strang splitting
of $\mathcal{R}_{i}(\{\bar{n}(t)\},\delta t)$ for one time step $\delta t$
separates the diffusion step and the injection step as follows 
\begin{align}
n_{i}^{*} & =n_{i}(t)+\mathcal{I}_{i}(\{\bar{n}(t)\},\delta t/2)\nonumber \\
n_{i}^{**} & =n_{i}^{*}+\mathcal{D}_{i}(\{\bar{n}^{*}\},\delta t)\nonumber \\
n_{i}(t+\delta t) & =n_{i}^{**}+\mathcal{I}_{i}(\{\bar{n}^{**}\},\delta t/2).\label{eq:StrangSplit}
\end{align}

\subsection{Diffusion step}

We would like that this method is extendable to arbitrary
reaction-diffusion particle-based simulations without spherical symmetry.
Therefore, we need to remove the constraint that the diffusion (and
reaction) in $r<R$ is modeled through jumps between spherical shells.
In order to do so, the diffusion step in the Strang splitting algorithm
can be done through a particle-based simulation, like over-damped
Langevin dynamics $dX_{t}=\sqrt{2D}dW_{t}$, with
$W_{t}$ a three-dimensional Wiener process, as commonly is the case
in reaction-diffusion particle-based simulations \citep{andrews2004stochastic,del2014fluorescence,schoneberg2013readdy},
see \citep{schoneberg2014simulation} for an overview. This can be
integrated with the Euler-Maruyama scheme \citep{higham2001algorithmic}
for each particle 
\begin{align}
X_{j}(t+\delta t)=X_{j}(t)+\sqrt{2D}\mathcal{N}(0,\delta t),\label{eq:EulerMaruyama}
\end{align}
where $D$ is the diffusion coefficient and $\mathcal{N}(0,\delta t)$
a three-dimensional vector with each entry a normal random variable
with mean zero and variance $\delta t$, and $j$ runs over all the
$B$ particles in the system. Note the number of $B$ particles is
not constant over time and that the diffusion coefficient
satisfies the Einstein relation, $D=k_{B}T\beta$, where $k_{B}$
is the Boltzmann constant, $T$ the temperature and $\beta$ a constant
related to the damping. If a particle diffuses into an absorbing
boundary (reaction) or into the region $r>R$, it is no longer considered
part of the system. We chose to use an absorbing
boundary (one absorbing A fixed at the origin) in order to verify
the numerical scheme with the analytic solution. However, as the reaction-diffusion
process is no longer restricted to spherical shells, the reaction-diffusion
particle-based simulation in $r<R$ can be arbitrary; we could have
chosen many diffusing reactive A's as well.

\subsection{Injection step}

In order to implement the injection of particles from the material
bath, it is convenient to adhere to the shell description. However,
it is only necessary to define two shells, the inner shell inside
the system $r\in(R-\delta r,R]$ and the outer shell $r\in(R,R+\delta r]$,
see Fig. \ref{fig:Gc_cartoon}a. The number of $B$ particles in the
outer shell $n_{c_{0}}$ has to be consistent with the far-field concentration
$c_{0}$, i.e we need to find reasonable value of $\delta r$ and
$n_{c_{0}}$ such that 
\begin{equation}
c_{0}=\frac{3n_{c_{0}}}{4\pi((R+\delta r)^{3}-R^{3})}.\label{eq:bathConc}
\end{equation}
As $\delta r$ is constrained by the time step of the simulation by
$2D\delta t\le\delta r^2$ \citep{del2016discrete}, we choose an initial
guess for $\delta r$ value slightly bigger than $\sqrt{2D\delta t}$. We
further choose $n_{c_{0}}$ as $4\pi c_{0}((R+\delta r)^{3}-R^{3})/3$
rounded up to the closest integer, and we do a Newton iteration on
Eq. (\ref{eq:bathConc}) as a function of $\delta r$. This will yield
the volume ($\delta r$) in which an integer number of particles yield
the concentration $c_{0}$ in the bath. It is important to check the
time-step constraint is fulfilled; if it is not, we require a higher
value for the initial guess of $\delta r$, which yields a larger
$n_{c_{0}}$ and consequently a larger $\delta r$. This method will
determine the number of ``virtual'' particles in the outer shell
(bath) and a consistent value for $\delta r$.

The ``virtual'' particles are injected from the bath (outer shell
$(R,R+\delta r]$) into the inner shell $(R-\delta r,R]$. Each particle
can jump inside with rate $\gamma(\delta r)$, as established in Eq.
(\ref{eq:gammmRate}). Note we write $\gamma(\delta r)$
to emphasize that $\gamma$ depends on the chosen $\delta r$. The
locations of the particles injected into the system are somewhere
within $(R-\delta r,R]$. As the exact location of the particles is
unknown due to the discrete resolution of the GC-SME, it is reasonable
to choose the location uniformly along $(R-\delta r,R]$. Once the
particles are injected into the system, they can diffuse following
the diffusion step.

\subsection{The particle-based scheme based on the GC-SME \label{subsec:PBscheme}}

The scheme described in Sec. \ref{sec:app_nummeth} is depicted in Fig.
\ref{fig:Gc_cartoon}a, and it is implemented as follows:

\texttt{\textbf{Input:}}\texttt{ Bath concentration $c_{0}$, diffusion
coefficient $D$, absorbing boundary $\sigma$, domain size $R$,
time-step $\delta_{t}$, initial guess for $\delta_{r}$, total number
of time iterations $m$ and Newton iteration tolerance $\epsilon$.} 
\begin{enumerate}
\item \texttt{Use Eq. (\ref{eq:bathConc}) to calculate $n_{c_{0}}$ (to
its nearest integer) and then approximate concentration $c_{0}^{\text{num}}$
for the given $\delta r$.} 
\item \texttt{While $|c_{0}^{\text{num}}-c_{0}|>\epsilon$:} 
\begin{enumerate}
\item \texttt{$\delta r$ $\leftarrow$ Newton iteration on Eq. (\ref{eq:bathConc}).} 
\item \texttt{Recalculate $c_{0}^{\text{num}}$.} 
\end{enumerate}
\item \texttt{If $2D\delta t>\delta r$:} 
\begin{enumerate}
\item \texttt{Exit, use larger initial guess for $\delta r$.} 
\end{enumerate}
\item \texttt{For $t=[0,\delta t,\dots,m\delta t]$:} 
\begin{enumerate}
\item \texttt{Inject particles from outer shell into inner shell for half
a time step with rate $\gamma(\delta r)$. Sample location of new
particles uniformly along $(R-\delta r,R]$.} 
\item \texttt{Diffuse all particles for $\delta t$ following the scheme
from Eq. (\ref{eq:EulerMaruyama}). If particles crossed the absorbing
boundary at $r=\sigma$ or the system domain at $r=R$, remove them.} 
\item \texttt{Inject particles from outer shell into inner shell for half
a time step with rate $\gamma(\delta r)$. Sample location of new
particles uniformly along $(R-\delta r,R]$.} 
\end{enumerate}
\end{enumerate}
Note steps 4a, 4b and 4c correspond to the Strang-splitting from Eq.
(\ref{eq:StrangSplit}). The results of simulations using this scheme
are shown in Fig. \ref{fig:LLN_conv}, where we further
verify the numerical scheme by showing the simulation results are
consistent with the analytic solution. These simulations were performed
using a purely absorbing reaction boundary. Nonetheless, it is straightforward
to extend the simulation to partially absorbing reaction boundaries
or more complicated reaction-diffusion schemes since any particle-based
scheme can be implemented in the diffusion step.

\section{Coupling particle and concentration-based simulations: an application
to exocytosis \label{sec:sim_exocyt} } 

In this section, we generalize the scheme from Sec.
\ref{sec:app_nummeth} by introducing a scheme to couple arbitrary
particle-based simulations with time and space-dependent bulk concentration
in the far-field. The far-field dynamics could be
known beforehand or obtained from another model, like reaction-diffusion
PDEs without high-order reactions. Interesting applications of this
scheme arise in intracellular biological processes triggered by changes
in calcium concentration through voltage-gated calcium channels. Examples
of such processes are secretion of endocrine and exocrine cells and
synaptic transmission, both of which occur through exocytosis \citep{bajjalieh1995biochemistry,barnes2002calcium,catterall2011voltage}.

Exocytosis consists of the active transport of molecules out of the
cell. The molecules to be transported are carried in vesicles towards
the cell membrane, where the vesicle fuses with the membrane expelling
all of its contents out of the cell. It is well known that exocytosis
is mainly triggered by changes in calcium concentration triggered
by voltage-gated calcium channels \citep{catterall2011voltage}. Considering
that vesicle diameters are around $50nm$ or larger \citep{bajjalieh1995biochemistry},
the domain of interest for a particle-based simulation should be at
least four times that size ($200nm$), which spans a volume of $8\times10^{6}nm^{3}$.
Calcium concentrations are very nonisotropic; in order to trigger
exocytosis, local calcium concentrations between $20\mu M$ and $1mM$
are required. This depends on the specific process, endocrine secretion
is around $27\mu M$ and synaptic transmission above $100\mu M$ \citep{bajjalieh1995biochemistry}.
This means that the number of particles in the volume of interest
could be between one hundred and several thousand, or even more if
the volume of interest is bigger. This number of particles is big
enough to roughly describe the calcium profile by a bulk concentration;
however, it is small enough that specific processes involving calcium
molecules would require a particle-based simulation. In other words,
it is ideal for a coupling like the one we are proposing.

\subsection{The model}

As a proof of concept, we begin with a simple model
inspired in exocytosis. We assume there is a set of $N_{S}$ calcium
sources in the cell membrane, each one of them located
at $x_{k}$ and producing a calcium concentration gradient $f_{k}(x,t)$
(with $k=1,2,\cdots,N_{s}$) in the intracellular space, Fig. \ref{fig:Gc_cartoon}b.
In the far-field, we are not interested in high-resolution dynamics,
so a concentration description in terms of a diffusion PDE is enough;
this could potentially be obtained from experiments. In
our case, the concentration gradient will simply be given by the sum
of solutions to diffusion PDEs with an initial concentration source
at $x_{k}$, 
\[
f_{k}(x,t)=\left(4\pi Dt\right)^{-3/2}\exp\left[-\frac{\left|x-x_{k}\right|^{2}}{4Dt}\right].
\]
We delimit the region of interest by half a sphere around the membrane,
Fig. \ref{fig:Gc_cartoon}b, where specific calcium-triggered process,
like exocytosis, could occur. In this region, we
can incorporate arbitrary particle-based reaction-diffusion simulations
to model processes like exocytosis in more detail; however, we restrict
ourselves to simple diffusion for verification purposes. Simulations
in this region could also include more detailed dynamics via MSM/RD
\citep{dibak2017msm}, where Markov state models (MSMs) \citep{bowman2014introduction,prinz2011markov,schutte2013metastability,trendelkamp2015estimation}
extracted from detailed molecular dynamics simulations are coupled
to particle-based reaction diffusion (RD) simulations. A more realistic
simulation of the membrane could also be implemented with the novel
particle-based membrane model from \citep{sadeghi2018particle}. We
should also note it is theoretically possible to do a particle-based
simulation or even a molecular dynamics simulation on the whole region.
In practice however this might be unfeasible, impractical and likely
unnecessary.

\begin{figure*}
\centering %
\noindent\begin{minipage}[c]{0.71\linewidth}%
 \centering \textbf{a.} \\
 \includegraphics[width=0.22\columnwidth]{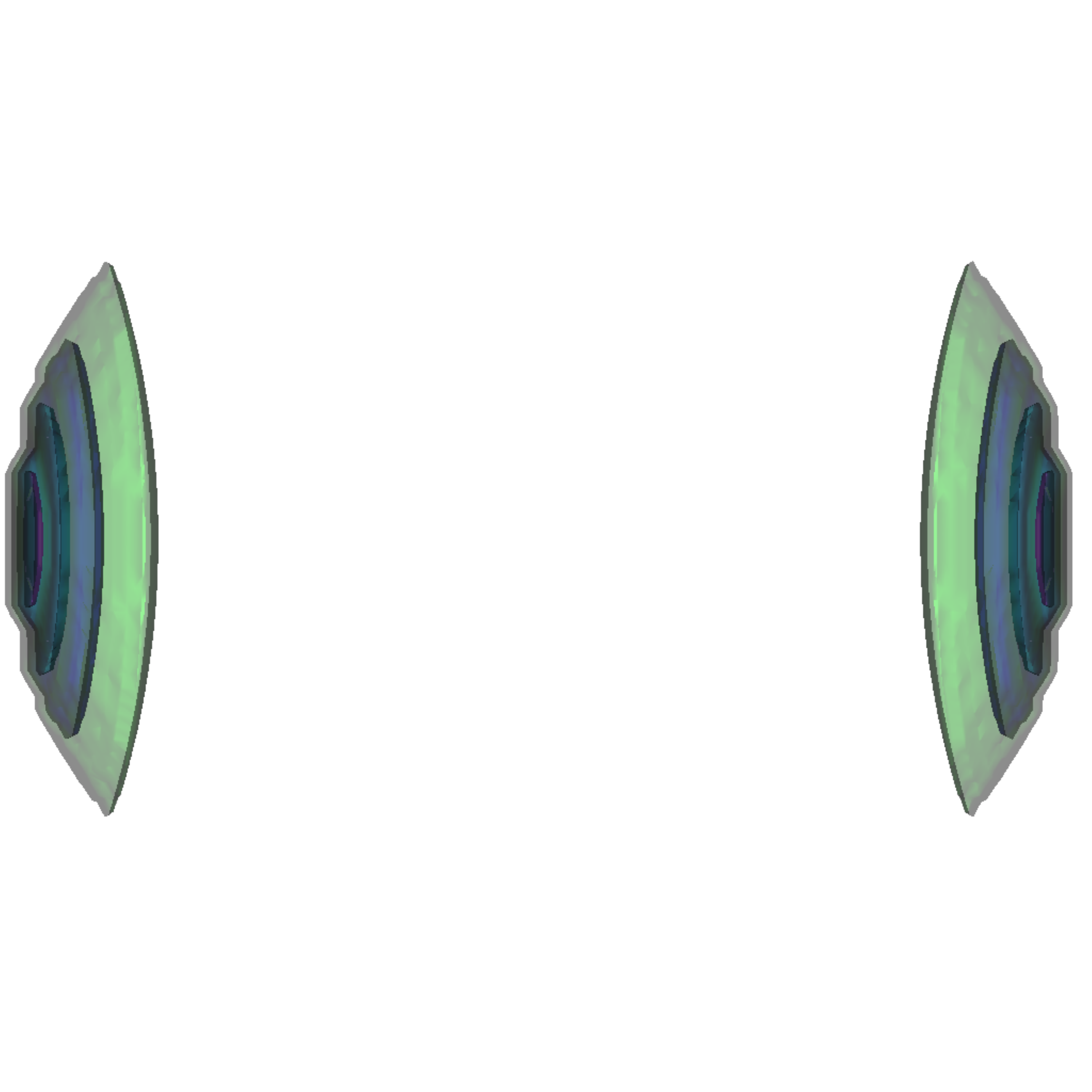}
\includegraphics[width=0.22\columnwidth]{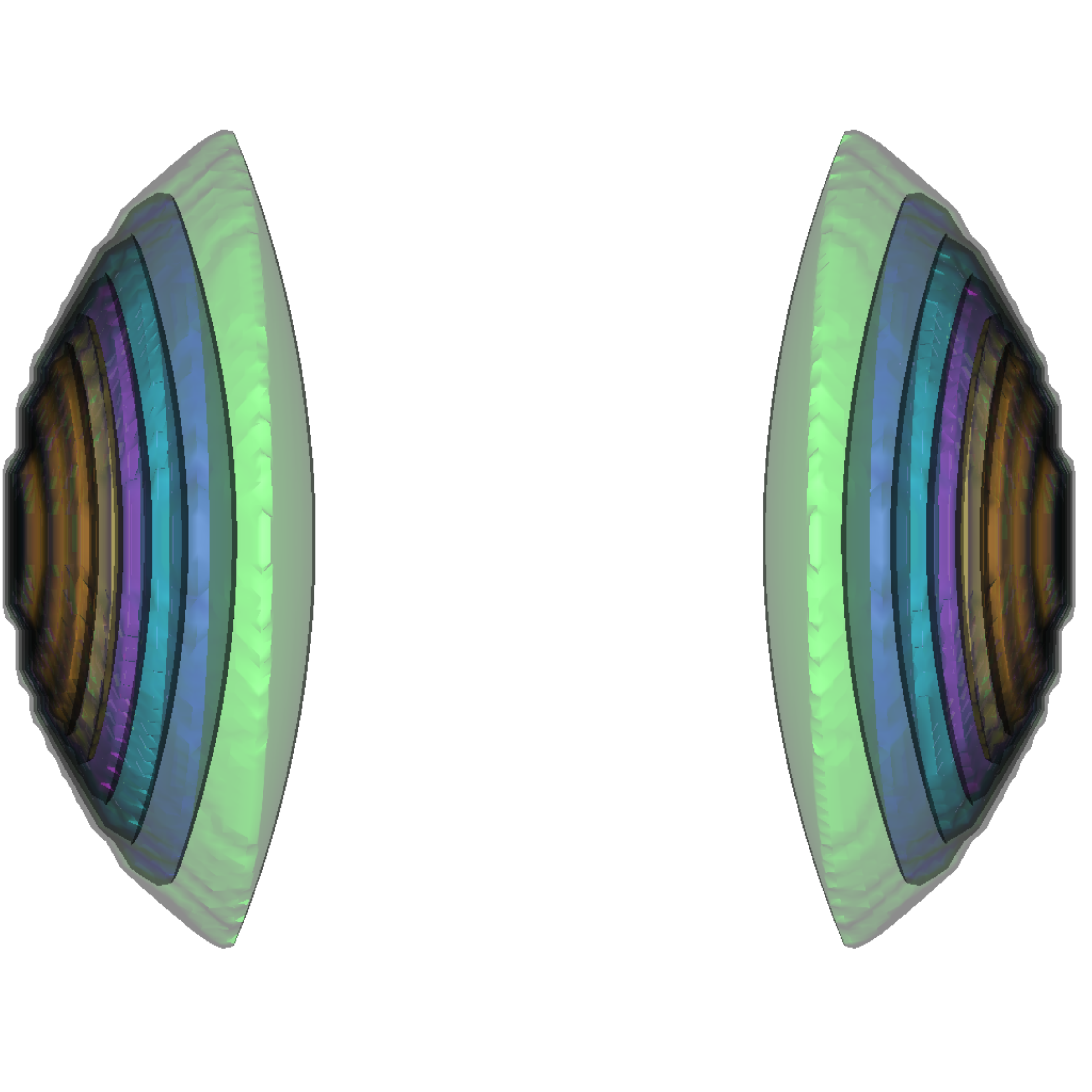}
\includegraphics[width=0.22\columnwidth]{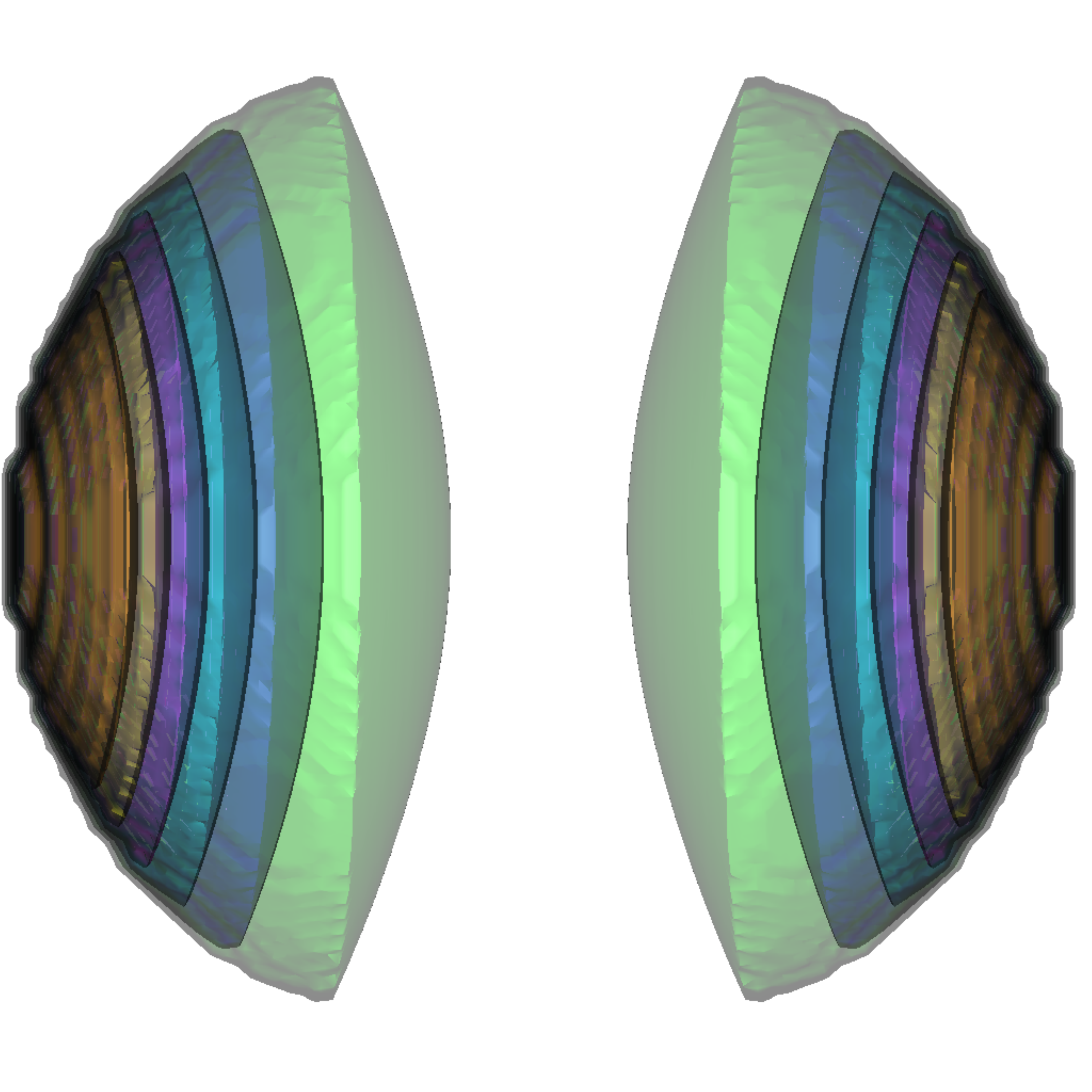}
\includegraphics[width=0.22\columnwidth]{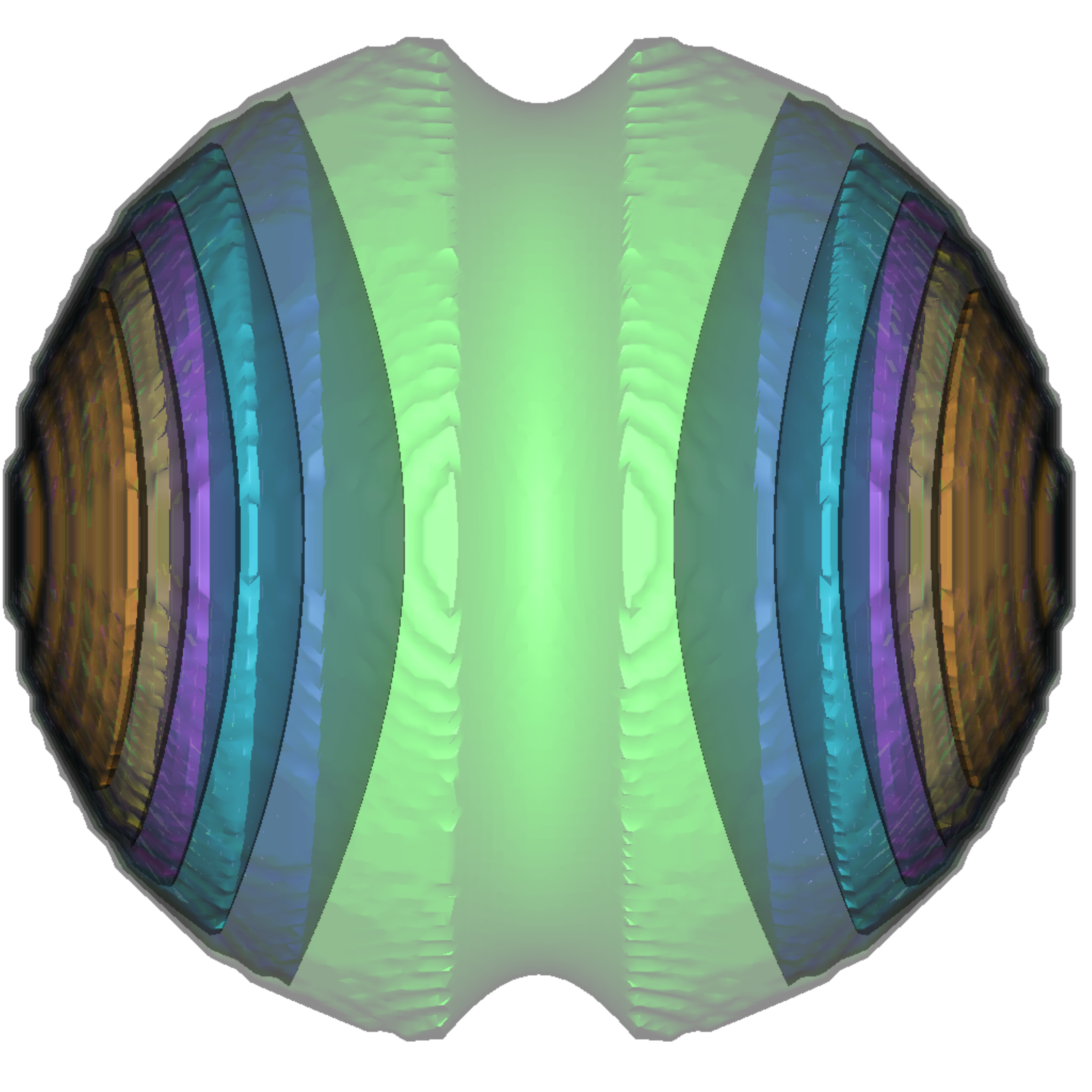}
\\
 \includegraphics[width=0.22\columnwidth]{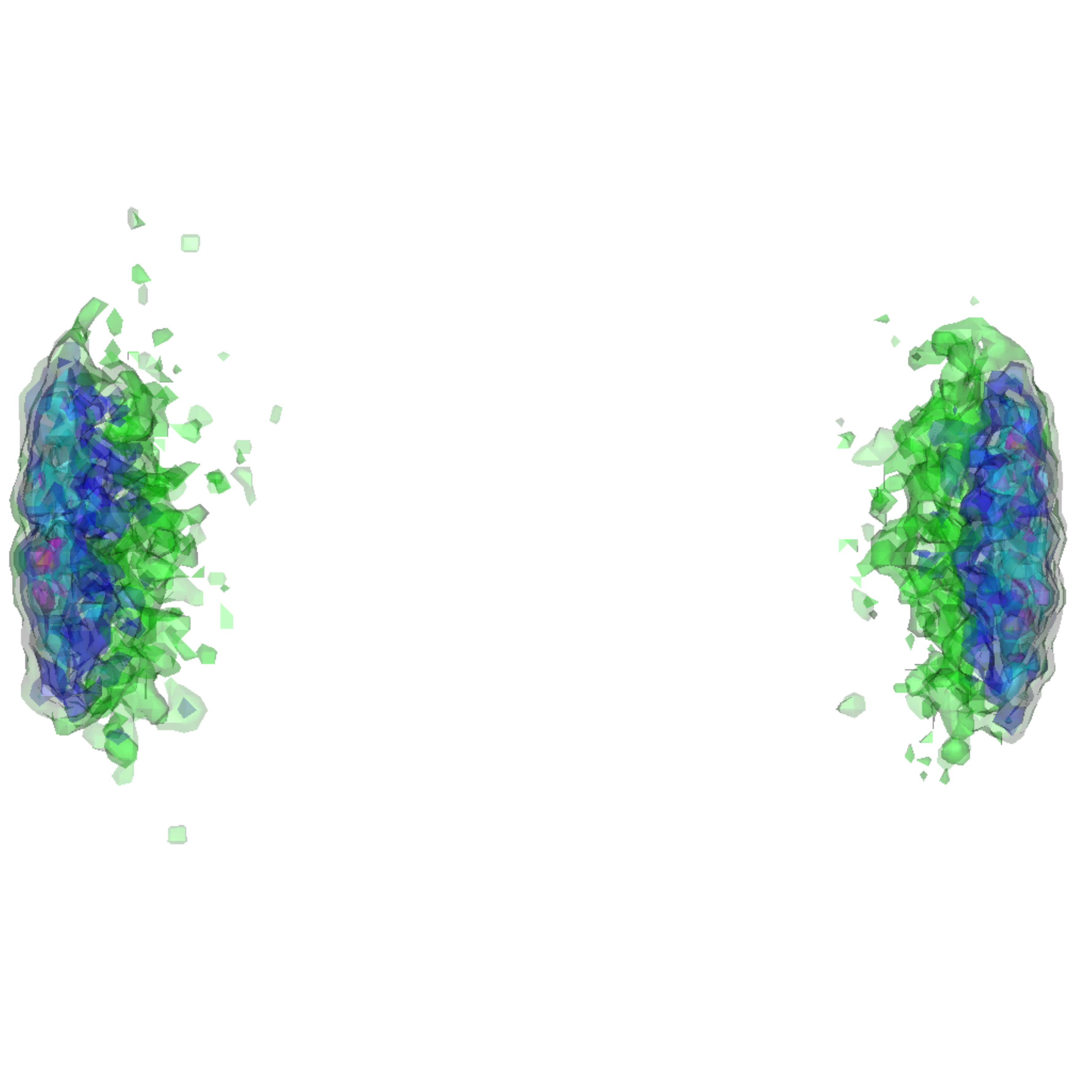}
\includegraphics[width=0.22\columnwidth]{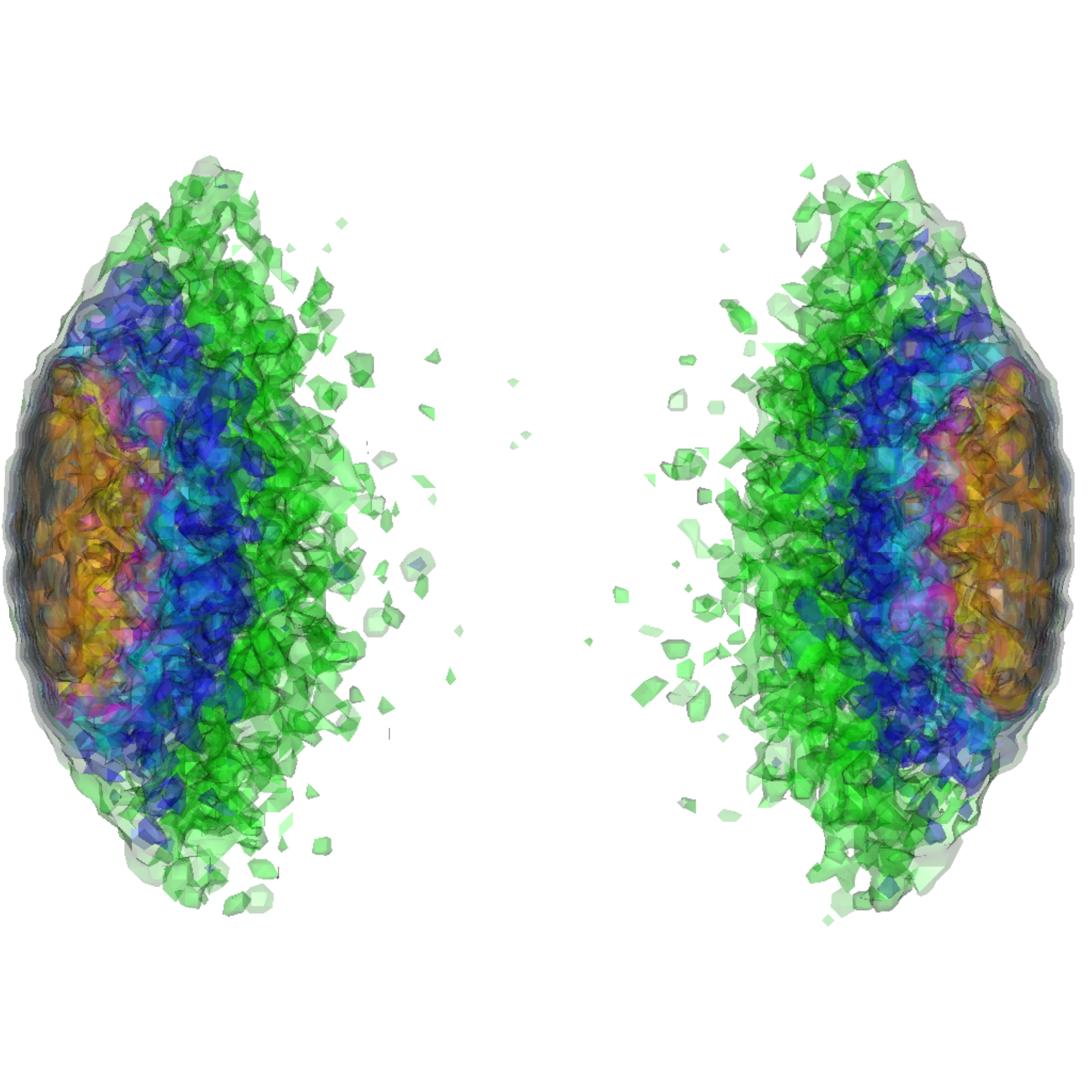}
\includegraphics[width=0.22\columnwidth]{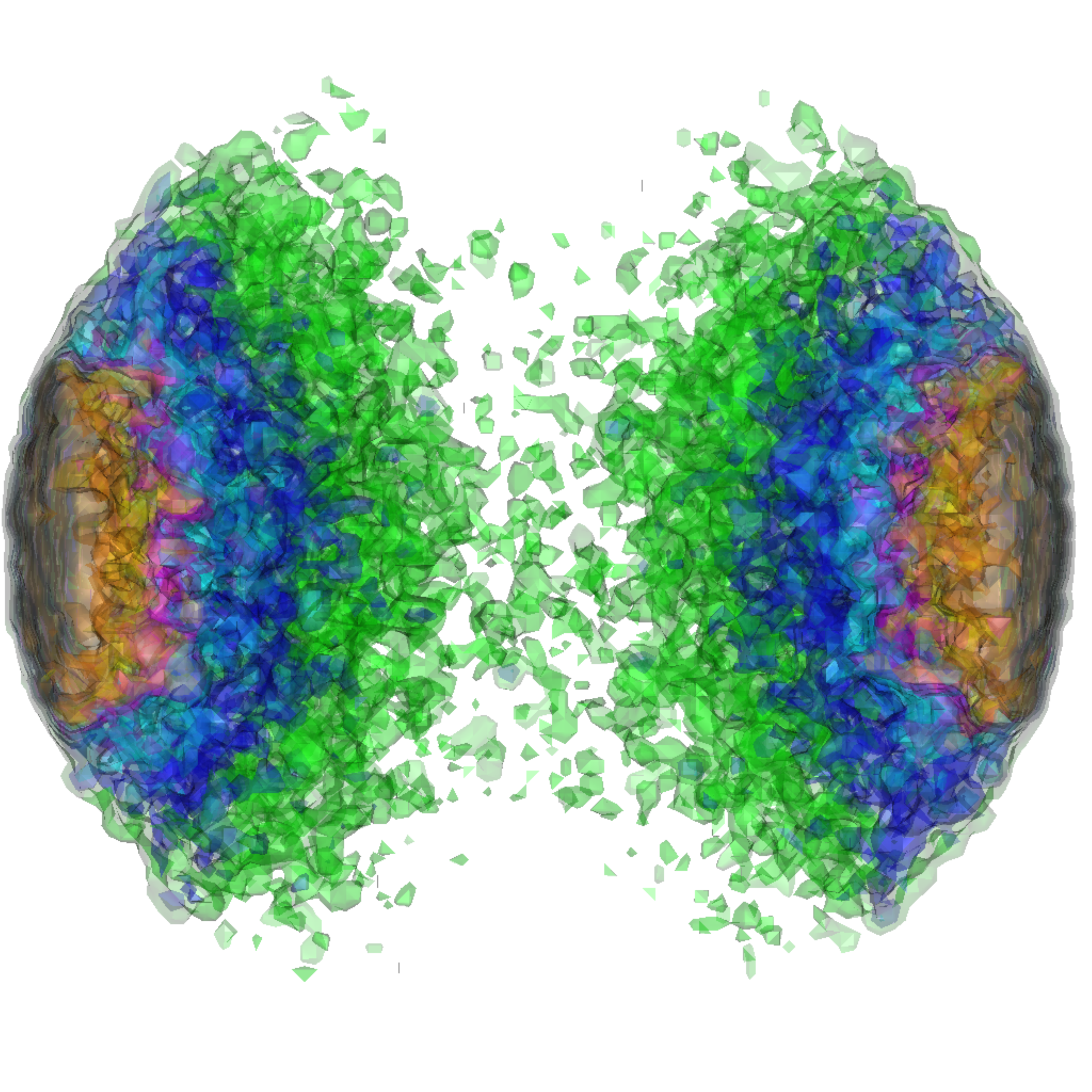}
\includegraphics[width=0.22\columnwidth]{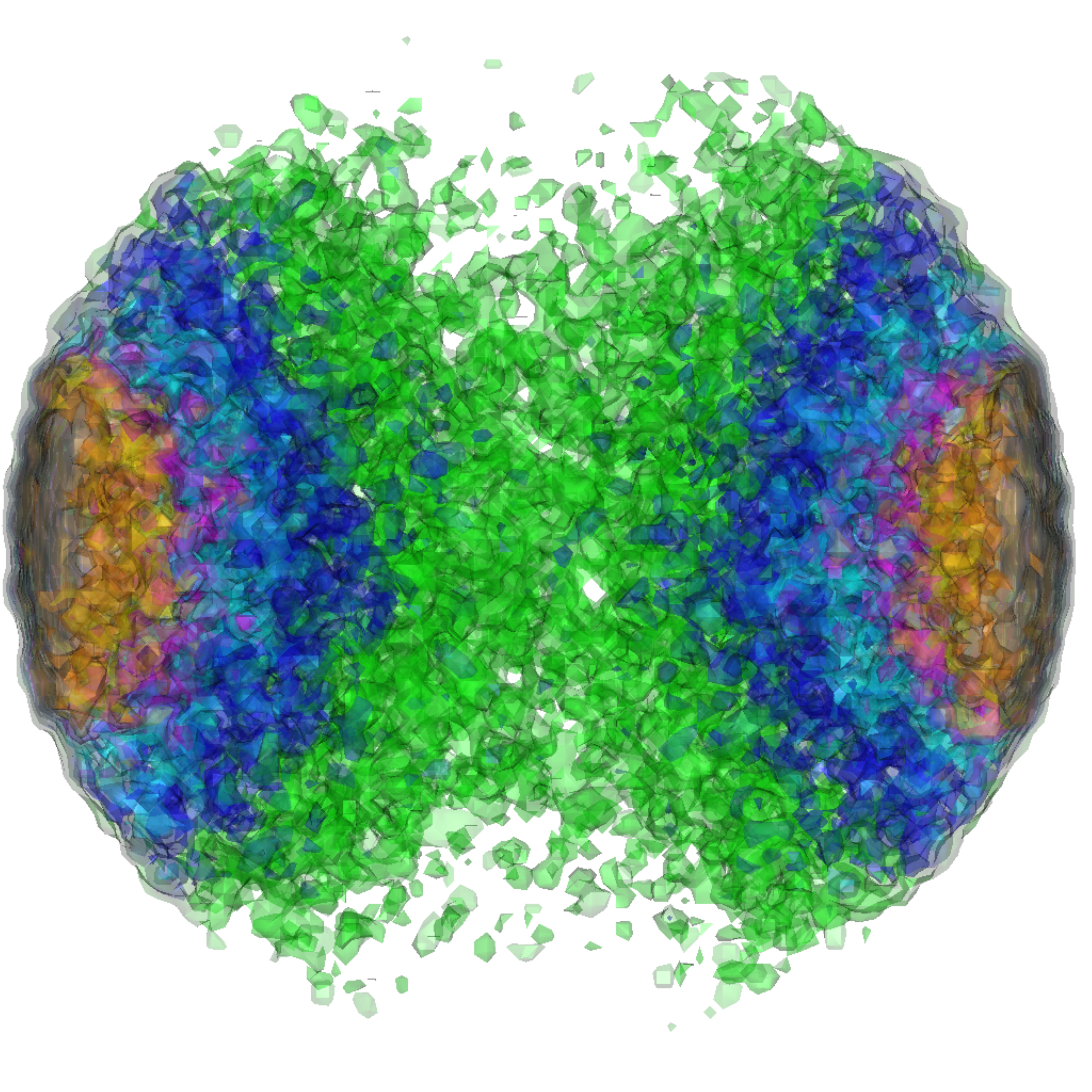} %
\end{minipage}%
\begin{minipage}[c]{0.27\linewidth}%
 \centering \textbf{b.} \includegraphics[width=1.1\columnwidth]{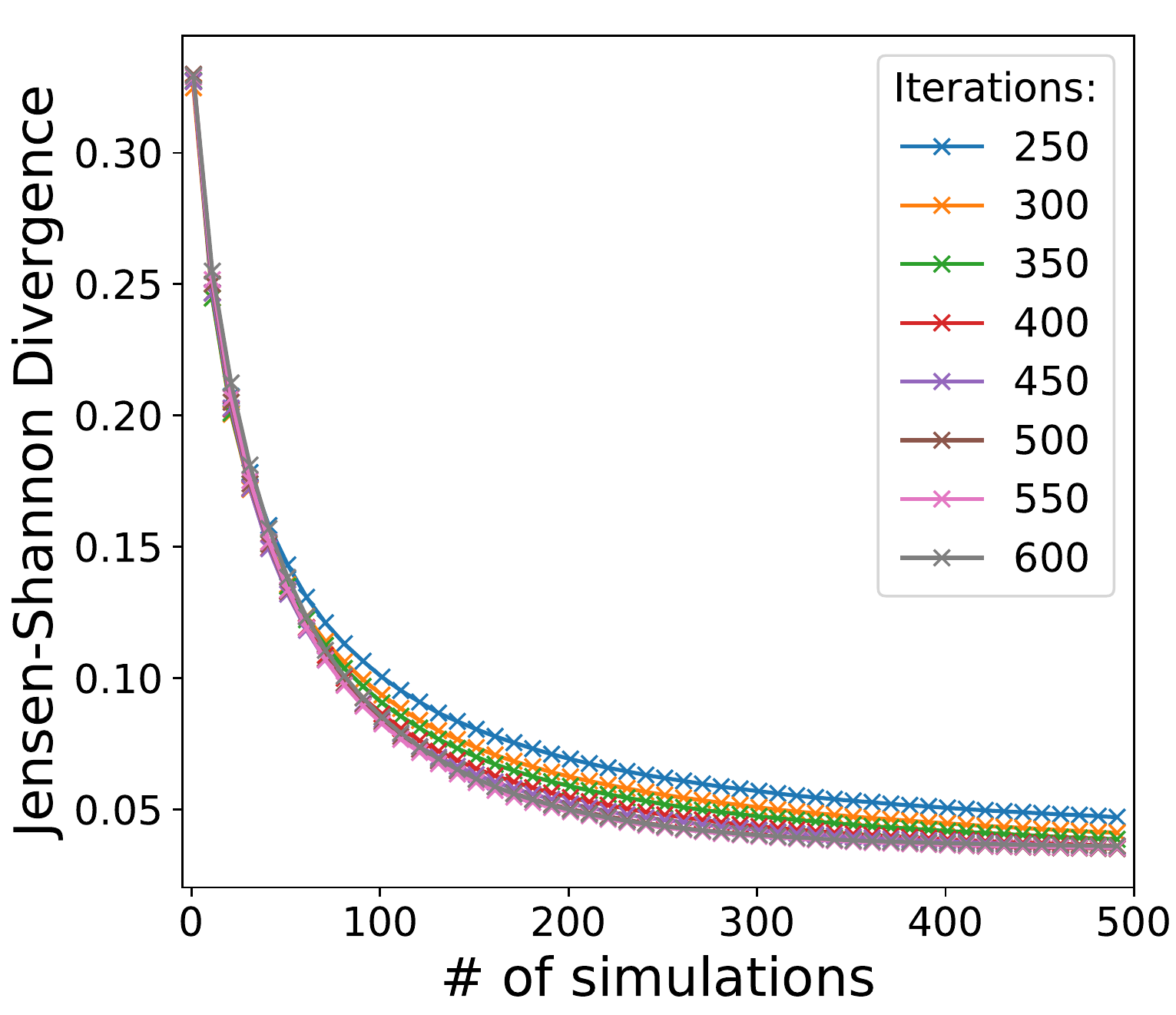}\\
\end{minipage}\caption{\textbf{a.} Contour plots of the Calcium concentration profile observed
from below the half sphere. The top plots correspond to the reference
based on bulk concentration dynamics while the bottom ones correspond
to the ensemble average of $200$ particle-based simulations. Both
simulations are plotted at four different points in time, $150,250,350$
and $450$ time iterations. Darker colors correspond to higher concentrations.
\textbf{b.} Plot of the Jensen-Shannon divergence between the reference
concentration histogram and the ensemble average
concentration histogram (over particle simulations) as a function
of the number of simulations used to calculate the ensemble average.
This curve is plotted to compare the histograms at different times
from $250$ to $650$ iterations. A value closer to zero means the
histograms are closer to each other. As more simulations are used
to calculate the average, we can observe the particle-based simulations
are in quantitative statistical agreement with the bulk concentration
dynamics.}
\label{fig:exocytContour} 
\end{figure*}

The main idea is to extend the model from Sec. \ref{sec:app_nummeth}
by incorporating angular and time resolution into the far-field concentration
value. In this case, we will also need an outer shell
covering the half sphere. However, we will need
to divide it into angular compartments, so we can emulate the space-dependent
material bath produced by the calcium concentration gradient. The
scheme is constructed in a similar manner to that of Sec. \ref{sec:app_nummeth},
but it is applied to each of the compartments. 

We begin by making an equal area partition of the surface of the half
sphere of radius $R$ into $p$ regions denoted by
$i=1,\cdots,p$ (Fig. \ref{fig:Gc_cartoon}c). We introduce two parameters
for each region, $\delta r_{i}$ and $n_{i}$. The first corresponds
to the width that generates a compartment $N_{i}$ of volume $\Omega_{i}$,
and the second corresponds to the number of particles inside that
compartment. Note each of the compartments, $N_{1},\cdots,N_{p}$,
has a different width $\delta r_{i}$, so it is delimited by $(R,R+\delta r_{i}]$
, Fig. \ref{fig:Gc_cartoon}c. Given a reasonable initial guess of
$\delta r_{i}$, we can estimate the volume of each compartment as
$2\pi R^{2}\delta r_{i}$ (an exact calculation is also possible),
and we can obtain a relation for the bulk concentration $c_{i}$ in
each compartment at time $t$ as 
\begin{equation}
c_{i}(t)=\frac{1}{\Omega_{i}}\int_{\omega_{i}}\left(\sum_{k=1}^{N_{S}}f_{k}(x,t)\right)d\omega_{i}.\label{eq:bathConcV2}
\end{equation}

With the implementation of the compartments, the
PDE solution for the far-field and this relation, we can derive the
numerical scheme. 

\subsection{Generalized particle-based scheme}

The scheme follows analogously from the Sec \ref{sec:app_nummeth}.
It will also be divided in a diffusion and injection
step; however, in order to model the time and space-dependent far-field
concentration dynamics, this algorithm needs to consider each of the
compartments, and requires the additional assumption that the bulk
concentration changes slowly over one time step $\delta t$. Note
that, although the diffusion step only implements particle diffusion,
it can be extended to arbitrary particle-based reaction-diffusion
simulations.

\texttt{\textbf{Input:}}\texttt{ Bath concentration $c_{0}$, diffusion
coefficient $D$, absorbing boundary $\sigma$, domain size $R$,
time-step $\delta_{t}$, initial guesses for $\delta r_{i}$, total
number of time iterations $m$, Newton iteration tolerance $\epsilon$,
partition chosen for compartments $\Omega_{i}$ and averaged bulk
concentration function in compartments $c_{i}(t)$.}

\texttt{For $t=[0,\delta t,\dots,m\delta t]$:} 
\begin{enumerate}
\item \texttt{For every compartment $i=1,\cdots,N_{p}$:} 
\begin{enumerate}
\item \texttt{Use Eq. (\ref{eq:bathConcV2}) to calculate $n_{i}$ (to its
nearest integer) and then approximate concentration $c_{i}(t)^{\text{num}}$
for the given $\delta r_{i}$.} 
\item \texttt{While $|c_{i}^{\text{num}}(t)-c_{i}(t)|>\epsilon$:} 
\begin{itemize}
\item \texttt{$\delta r_{i}$ $\leftarrow$ Newton iteration on Eq. (\ref{eq:bathConcV2}).} 
\item \texttt{Recalculate $c_{i}^{\text{num}}(t)$.} 
\end{itemize}
\item \texttt{If $2D\delta t>\delta r_{i}$:} 
\begin{itemize}
\item \texttt{Use a larger initial guess for $\delta r_{i}$ and repeat
1.a and 1.b for current compartment.} 
\end{itemize}
\end{enumerate}
\item \texttt{For every compartment $i=1,\cdots,N_{p}$:} 
\begin{enumerate}
\item \texttt{Inject particles from compartment $i$ into system for half
a time step with rate $\gamma_{i}(\delta r_{i})$. Sample location
of new particles uniformly in the region delimited by the partition
$i$ and $(R-\delta r_{i},R]$.} 
\end{enumerate}
\item \texttt{Diffuse all particles for $\delta t$ following the scheme
from Eq. (\ref{eq:EulerMaruyama}). If particles crossed the system
domain at $r=R$, remove them.} 
\item \texttt{For every compartment $i=1,\cdots,N_{p}$:} 
\begin{enumerate}
\item \texttt{Inject particles from compartment $i$ into system for half
a time step with rate $\gamma_{i}(\delta r_{i})$. Sample location
of new particles uniformly in the region delimited by the partition
$i$ and $(R-\delta r_{i},R]$.} 
\end{enumerate}
\end{enumerate}
Following the setup from Fig. \ref{fig:Gc_cartoon}b, we illustrate
the results for a ''proof of concept'' simulation with two sources
for the Calcium concentration at distances $\lambda_{1}=\lambda_{2}=5$
from the origin. Fig. \ref{fig:exocytContour}a shows the concentration
contour plots in the half-sphere region (with radius $R=5$) (seen
from below) for an average of $200$ particle-based simulation along
with the reference bulk concentration solution at four different times.
Furthermore, we compared the three-dimensional concentration histograms
obtained from the particle-based simulations ensemble averages against
the reference concentration solutions. The similarity
between the two distributions was measured by calculating the Jensen-Shannon
divergence \citep{lin1991divergence} between the two normalized concentration
histograms; Fig. \ref{fig:exocytContour}b shows this comparison
as a function of the number of simulations used to calculate the ensemble
average at different points in time. We observe the particle-based
simulation are in quantitative statistical agreement with the bulk
concentration dynamics, which validates the simulation results. Although
we constructed this scheme with exocytosis applications in mind, it
could be implemented for many other systems.

Note it is straightforward to extend the bulk concentration description
to include several species and unimolecular reactions, where the dynamics
are in terms of reaction-diffusion PDEs restricted to first-order
reactions. Extending the scheme to incorporate higher-order reactions
in the coupling boundary is not trivial since higher-order reactions
are no longer independent of diffusion. This is also an issue in \citep{franz2013multiscale};
however, we think the GC-SME provides a robust framework that will
be helpful to address this issue in future work. Nonetheless, the
current scheme can be generalized to arbitrarily complicated systems
in the particle-based region. 

\section{Conclusion \label{sec:conclusions}}

We constructed continuous-time discrete-state Markov
models for diffusion-influenced reactions (SMEs). The first SME corresponds
to the case of an isolated pair of reacting particles. In the continuous
limit, it recovers Smoluchowski's probabilistic approach. We later
introduced the GC-SME, a generalization of the previous SME to an
arbitrary non-constant number of ligand particles. In the continuous
limit, when taking either the hydrodynamic or the large copy number
limit, the GC-SME converges to Smoluchowski's concentration-based
approach with a constant concentration in the far-field. We finally
employed this result to implement two particle-based simulations coupled
to bulk concentration descriptions.

The GC-SME convergence result addresses several
matters of relevance to the theory of diffusion-limited reactions
and stochastic reaction-diffusion processes. First of all, it establishes
a precise connection between the probabilistic and concentration-based
approach as well as an interpretation of the concentration-based approach
in terms of a probabilistic model; this is essentially an extension
of Kurtz limit \citep{kurtz1971limit,kurtz1972relationship} to a
class of spatially inhomogeneous chemical systems. In addition, it
provides a robust framework for statistical mechanical interpretation,
which clarifies interpretations at the particle level and bridges
the concept of chemical potential from a mesoscopic to a macroscopic
scale. In a more pragmatic note, it enables multiscale and hybrid
particle-based schemes by consistently coupling them to reaction-diffusion
PDEs (with only first-order reactions).

The results in this paper provide the blueprints
for multiscale/hybrid numerical frameworks that could potentially
couple particle-based reaction-diffusion simulations with general
reaction-diffusion PDEs. However, our current approach only allows
for higher-order reactions to occur inside the particle-based domain
since it can only couple diffusion processes across the particle-based
simulation boundary. Therefore, it is not yet clear how can one couple
high-order reaction processes consistently. We leave this endeavor
for future work.

\section{Acknowledgments \label{sec:ack} }

We gratefully acknowledge support by the Deutsche Forschungsgemeinschaft
(grants SFB1114, projects C03 and A04), the Einstein Foundation Berlin
(ECMath grant CH17), the European research council (ERC starting grant
307494 ``pcCell'') and the National Science and Technology Council
of Mexico (CONACYT). We also thank Attila Szabo for insightful discussions
and encouragements over the course of this work.

\bibliographystyle{IEEEtran}
\bibliography{prob_smol_refs}

\begin{thebibliography}{10}
\providecommand{\url}[1]{#1}
\csname url@samestyle\endcsname
\providecommand{\newblock}{\relax}
\providecommand{\bibinfo}[2]{#2}
\providecommand{\BIBentrySTDinterwordspacing}{\spaceskip=0pt\relax}
\providecommand{\BIBentryALTinterwordstretchfactor}{4}
\providecommand{\BIBentryALTinterwordspacing}{\spaceskip=\fontdimen2\font plus
\BIBentryALTinterwordstretchfactor\fontdimen3\font minus
  \fontdimen4\font\relax}
\providecommand{\BIBforeignlanguage}[2]{{%
\expandafter\ifx\csname l@#1\endcsname\relax
\typeout{** WARNING: IEEEtran.bst: No hyphenation pattern has been}%
\typeout{** loaded for the language `#1'. Using the pattern for}%
\typeout{** the default language instead.}%
\else
\language=\csname l@#1\endcsname
\fi
#2}}
\providecommand{\BIBdecl}{\relax}
\BIBdecl

\bibitem{collins1949diffusion}
F.~C. Collins and G.~E. Kimball, ``Diffusion-controlled reaction rates,''
  \emph{J. Colloid Sci.}, vol.~4, no.~4, pp. 425--437, 1949.

\bibitem{smoluchowski1917versuch}
M.~von Smoluchowski, ``Versuch einer mathematischen theorie der
  koagulationskinetik kolloider l{\"o}sungen,'' \emph{Z. Phys. Chem}, vol.~92,
  no. 129-168, p.~9, 1917.

\bibitem{agmon1990theory}
N.~Agmon and A.~Szabo, ``Theory of reversible diffusion-influenced reactions,''
  \emph{J. Chem. Phys.}, vol.~92, no.~9, pp. 5270--5284, 1990.

\bibitem{hanggi1990reaction}
P.~H{\"a}nggi, P.~Talkner, and M.~Borkovec, ``Reaction-rate theory: Fifty years
  after {K}ramers,'' \emph{Rev. Mod. Phys.}, vol.~62, no.~2, p. 251, 1990.

\bibitem{keizer1982nonequilibrium}
J.~Keizer, ``Nonequilibrium statistical thermodynamics and the effect of
  diffusion on chemical reaction rates,'' \emph{The Journal of Physical
  Chemistry}, vol.~86, no.~26, pp. 5052--5067, 1982.

\bibitem{keizer1987diffusion}
------, ``Diffusion effects on rapid bimolecular chemical reactions,''
  \emph{Chemical Reviews}, vol.~87, no.~1, pp. 167--180, 1987.

\bibitem{schurr-1970}
J.~M. Schurr, ``The role of diffusion in bimolecular solution kinetics,''
  \emph{Biophys. J.}, vol.~10, no.~8, p. 700, 1970.

\bibitem{szabo1980first}
A.~Szabo, K.~Schulten, and Z.~Schulten, ``First passage time approach to
  diffusion controlled reactions,'' \emph{J. Chem. Phys.}, vol.~72, no.~8, pp.
  4350--4357, 1980.

\bibitem{szabo1989theory}
A.~Szabo, ``Theory of diffusion-influenced fluorescence quenching,'' \emph{J.
  Phys. Chem.}, vol.~93, no.~19, pp. 6929--6939, 1989.

\bibitem{del2014fluorescence}
M.~J. Del~Razo, W.~Pan, H.~Qian, and G.~Lin, ``Fluorescence correlation
  spectroscopy and nonlinear stochastic reaction--diffusion,'' \emph{The
  Journal of Physical Chemistry B}, vol. 118, no.~25, pp. 7037--7046, 2014.

\bibitem{del2016discrete}
M.~J. del Razo and H.~Qian, ``A discrete stochastic formulation for reversible
  bimolecular reactions via diffusion encounter,'' \emph{Comm. Math. Sci.},
  vol.~14, no.~6, pp. 1741--1772, 2016.

\bibitem{dorsaz2010diffusion}
N.~Dorsaz, C.~De~Michele, F.~Piazza, P.~De~Los~Rios, and G.~Foffi,
  ``Diffusion-limited reactions in crowded environments,'' \emph{Phys. Rev.
  Lett.}, vol. 105, no.~12, p. 120601, 2010.

\bibitem{donev2010first}
A.~Donev, V.~V. Bulatov, T.~Oppelstrup, G.~H. Gilmer, B.~Sadigh, and M.~H.
  Kalos, ``A first-passage kinetic monte carlo algorithm for complex
  diffusion--reaction systems,'' \emph{J. Comput. Phys.}, vol. 229, no.~9, pp.
  3214--3236, 2010.

\bibitem{gopich2002kinetics}
I.~V. Gopich and A.~v. Szabo, ``Kinetics of reversible diffusion influenced
  reactions: the self-consistent relaxation time approximation,'' \emph{J.
  Chem. Phys.}, vol. 117, no.~2, pp. 507--517, 2002.

\bibitem{hagen1996diffusion}
S.~J. Hagen, J.~Hofrichter, A.~Szabo, and W.~A. Eaton, ``Diffusion-limited
  contact formation in unfolded cytochrome c: estimating the maximum rate of
  protein folding,'' \emph{Proc. Natl. Acad. Sci. U.S.A.}, vol.~93, no.~21, pp.
  11\,615--11\,617, 1996.

\bibitem{peters2013reaction}
B.~Peters, P.~G. Bolhuis, R.~G. Mullen, and J.-E. Shea, ``Reaction coordinates,
  one-dimensional {S}moluchowski equations, and a test for dynamical
  self-consistency,'' \emph{J. Chem. Phys.}, vol. 138, no.~5, p. 054106, 2013.

\bibitem{prustel2013theory}
T.~Pr{\"u}stel and M.~Meier-Schellersheim, ``Theory of reversible
  diffusion-influenced reactions with non-{M}arkovian dissociation in two space
  dimensions,'' \emph{J. Chem. Phys.}, vol. 138, no.~10, p. 104112, 2013.

\bibitem{tucci2004mesoscopic}
K.~Tucci and R.~Kapral, ``Mesoscopic model for diffusion-influenced reaction
  dynamics,'' \emph{J. Chem. Phys.}, vol. 120, no.~17, pp. 8262--8270, 2004.

\bibitem{vijaykumar2015combining}
A.~Vijaykumar, P.~G. Bolhuis, and P.~R. ten Wolde, ``Combining molecular
  dynamics with mesoscopic {G}reen's function reaction dynamics simulations,''
  \emph{J. Chem. Phys.}, vol. 143, no.~21, p. 214102, 2015.

\bibitem{vijaykumar2017multiscale}
A.~Vijaykumar, T.~E. Ouldridge, P.~R. ten Wolde, and P.~G. Bolhuis,
  ``Multiscale simulations of anisotropic particles combining molecular
  dynamics and {G}reen's function reaction dynamics,'' \emph{J. Chem. Phys.},
  vol. 146, no.~11, p. 114106, 2017.

\bibitem{van2005green}
J.~S. van Zon and P.~R. Ten~Wolde, ``Green's-function reaction dynamics: A
  particle-based approach for simulating biochemical networks in time and
  space,'' \emph{J. Chem. Phys.}, vol. 123, no.~23, p. 4910, 2005.

\bibitem{szabo2008autobiography}
A.~Szabo, ``Autobiography of {A}ttila {S}zabo,'' \emph{J. Phys. Chem. B}, vol.
  112, no.~19, pp. 5883--5891, 2008.

\bibitem{berg1978diffusion}
O.~G. Berg, ``On diffusion-controlled dissociation,'' \emph{Chem. Phys.},
  vol.~31, no.~1, pp. 47--57, 1978.

\bibitem{gopich2002asymptotic}
I.~V. Gopich and A.~Szabo, ``Asymptotic relaxation of reversible bimolecular
  chemical reactions,'' \emph{Chem. Phys.}, vol. 284, no. 1-2, pp. 91--102,
  2002.

\bibitem{franz2013multiscale}
B.~Franz, M.~B. Flegg, S.~J. Chapman, and R.~Erban, ``Multiscale
  reaction-diffusion algorithms: {PDE}-assisted {B}rownian dynamics,''
  \emph{SIAM J. Appl. Math.}, vol.~73, no.~3, pp. 1224--1247, 2013.

\bibitem{smith2018spatial}
S.~Smith and R.~Grima, ``Spatial stochastic intracellular kinetics: A review of
  modelling approaches,'' \emph{Bull. Math. Biol.}, pp. 1--50, 2018.

\bibitem{doi1976stochastic}
M.~Doi, ``Stochastic theory of diffusion-controlled reaction,'' \emph{J. Phys.
  A: Math. Gen.}, vol.~9, no.~9, p. 1479, 1976.

\bibitem{erban2009stochastic}
R.~Erban and S.~J. Chapman, ``Stochastic modelling of reaction--diffusion
  processes: algorithms for bimolecular reactions,'' \emph{Phys. Biol.},
  vol.~6, no.~4, p. 046001, 2009.

\bibitem{isaacson2009reaction}
S.~A. Isaacson, ``The reaction-diffusion master equation as an asymptotic
  approximation of diffusion to a small target,'' \emph{SIAM J. Appl. Math.},
  vol.~70, no.~1, pp. 77--111, 2009.

\bibitem{isaacson2013convergent}
------, ``A convergent reaction-diffusion master equation,'' \emph{J. Chem.
  Phys.}, vol. 139, no.~5, p. 054101, 2013.

\bibitem{franco2014interacting}
T.~Franco, ``Interacting particle systems: hydrodynamic limit versus high
  density limit,'' in \emph{From Particle Systems to Partial Differential
  Equations}.\hskip 1em plus 0.5em minus 0.4em\relax Springer, 2014, pp.
  179--189.

\bibitem{kipnis2013scaling}
C.~Kipnis and C.~Landim, \emph{Scaling limits of interacting particle
  systems}.\hskip 1em plus 0.5em minus 0.4em\relax Springer Science \& Business
  Media, 2013, vol. 320.

\bibitem{kurtz1971limit}
T.~G. Kurtz, ``Limit theorems for sequences of jump {M}arkov processes
  approximating ordinary differential processes,'' \emph{J. Appl. Probab.},
  vol.~8, no.~2, pp. 344--356, 1971.

\bibitem{kurtz1972relationship}
T.~G.~v. Kurtz, ``The relationship between stochastic and deterministic models
  for chemical reactions,'' \emph{J. Chem. Phys.}, vol.~57, no.~7, pp.
  2976--2978, 1972.

\bibitem{qian2010chemical}
H.~Qian and L.~M. Bishop, ``The chemical master equation approach to
  nonequilibrium steady-state of open biochemical systems: linear
  single-molecule enzyme kinetics and nonlinear biochemical reaction
  networks,'' \emph{Int. J. Mol. Sci.}, vol.~11, no.~9, pp. 3472--3500, 2010.

\bibitem{corry2000tests}
B.~Corry, S.~Kuyucak, and S.-H. Chung, ``Tests of continuum theories as models
  of ion channels. ii. poisson--nernst--planck theory versus {B}rownian
  dynamics,'' \emph{Biophys. J.}, vol.~78, no.~5, pp. 2364--2381, 2000.

\bibitem{pedersen2011mathematical}
M.~G. Pedersen, G.~Cortese, and L.~Eliasson, ``Mathematical modeling and
  statistical analysis of calcium-regulated insulin granule exocytosis in
  $\beta$-cells from mice and humans,'' \emph{Prog. Biophys. Mol. Biol.}, vol.
  107, no.~2, pp. 257--264, 2011.

\bibitem{zhuravlev2009molecular}
P.~I. Zhuravlev and G.~A. Papoian, ``Molecular noise of capping protein binding
  induces macroscopic instability in filopodial dynamics,'' \emph{Proc. Natl.
  Acad. Sci. U.S.A.}, vol. 106, no.~28, pp. 11\,570--11\,575, 2009.

\bibitem{heuett2006grand}
W.~J. Heuett and H.~Qian, ``Grand canonical {M}arkov model: a stochastic theory
  for open nonequilibrium biochemical networks,'' \emph{J. Chem. Phys.}, vol.
  124, no.~4, p. 044110, 2006.

\bibitem{shoup1982role}
D.~Shoup and A.~Szabo, ``Role of diffusion in ligand binding to macromolecules
  and cell-bound receptors,'' \emph{Biophys. J.}, vol.~40, no.~1, pp. 33--39,
  1982.

\bibitem{sokolowski2010green}
T.~Sokolowski, L.~Bossen, T.~Miedema, and N.~Becker, ``Green's function
  reaction dynamics--an exact and efficient way to simulate intracellular
  pattern formation,'' in \emph{ICNAAM 2010}, vol. 1281, no.~1.\hskip 1em plus
  0.5em minus 0.4em\relax AIP Publishing, 2010, pp. 1342--1345.

\bibitem{van2005simulating}
J.~S. van Zon and P.~R. Ten~Wolde, ``Simulating biochemical networks at the
  particle level and in time and space: {G}reen's function reaction dynamics,''
  \emph{Phys. Rev. Lett.}, vol.~94, no.~12, p. 128103, 2005.

\bibitem{ye2018dynamic}
F.~X.-F. Ye, P.~Stinis, and H.~Qian, ``Dynamic looping of a free-draining
  polymer,'' \emph{SIAM J. Appl. Math.}, vol.~78, no.~1, pp. 104--123, 2018.

\bibitem{wang2003robust}
H.~Wang, C.~S. Peskin, and T.~C. Elston, ``A robust numerical algorithm for
  studying biomolecular transport processes,'' \emph{J. Theor. Biol.}, vol.
  221, no.~4, pp. 491--511, 2003.

\bibitem{pathria1996statistical}
R.~Pathria and P.~Beale, \emph{Statistical Mechanics}.\hskip 1em plus 0.5em
  minus 0.4em\relax Elsevier Science, 1996.

\bibitem{bedeaux2010mesoscopic}
D.~Bedeaux, I.~Pagonabarraga, J.~O. De~Z{\'a}rate, J.~Sengers, and
  S.~Kjelstrup, ``Mesoscopic non-equilibrium thermodynamics of non-isothermal
  reaction-diffusion,'' \emph{Phys. Chem. Chem. Phys.}, vol.~12, no.~39, pp.
  12\,780--12\,793, 2010.

\bibitem{ge2016mesoscopic}
H.~Ge and H.~Qian, ``Mesoscopic kinetic basis of macroscopic chemical
  thermodynamics: A mathematical theory,'' \emph{Phys. Rev. E}, vol.~94, no.~5,
  p. 052150, 2016.

\bibitem{ge2017mathematical}
H.~Ge and H.~v. Qian, ``Mathematical formalism of nonequilibrium thermodynamics
  for nonlinear chemical reaction systems with general rate law,'' \emph{J.
  Stat. Phys.}, vol. 166, no.~1, pp. 190--209, 2017.

\bibitem{anderson2015stochastic}
D.~F. Anderson and T.~G. Kurtz, \emph{Stochastic analysis of biochemical
  systems}.\hskip 1em plus 0.5em minus 0.4em\relax Springer, 2015, vol.~1.

\bibitem{kim2017stochastic}
C.~Kim, A.~Nonaka, J.~B. Bell, A.~L. Garcia, and A.~Donev, ``Stochastic
  simulation of reaction-diffusion systems: A fluctuating-hydrodynamics
  approach,'' \emph{J. Chem. Phys.}, vol. 146, no.~12, p. 124110, 2017.

\bibitem{andrews2004stochastic}
S.~S. Andrews and D.~Bray, ``Stochastic simulation of chemical reactions with
  spatial resolution and single molecule detail,'' \emph{Phys. Biol.}, vol.~1,
  no.~3, p. 137, 2004.

\bibitem{schoneberg2013readdy}
J.~Sch{\"o}neberg and F.~No{\'e}, ``Readdy-a software for particle-based
  reaction-diffusion dynamics in crowded cellular environments,'' \emph{PloS
  one}, vol.~8, no.~9, p. e74261, 2013.

\bibitem{schoneberg2014simulation}
J.~Sch{\"o}neberg, A.~Ullrich, and F.~No{\'e}, ``Simulation tools for
  particle-based reaction-diffusion dynamics in continuous space,'' \emph{BMC
  biophysics}, vol.~7, no.~1, p.~11, 2014.

\bibitem{higham2001algorithmic}
D.~J. Higham, ``An algorithmic introduction to numerical simulation of
  stochastic differential equations,'' \emph{SIAM review}, vol.~43, no.~3, pp.
  525--546, 2001.

\bibitem{bajjalieh1995biochemistry}
S.~M. Bajjalieh and R.~H. Scheller, ``The biochemistry of neurotransmitter
  secretion,'' \emph{J. Biol. Chem.}, vol. 270, no.~5, pp. 1971--1974, 1995.

\bibitem{barnes2002calcium}
S.~Barnes and M.~E. Kelly, ``Calcium channels at the photoreceptor synapse,''
  in \emph{Photoreceptors and calcium}.\hskip 1em plus 0.5em minus 0.4em\relax
  Springer, 2002, pp. 465--476.

\bibitem{catterall2011voltage}
W.~A. Catterall, ``Voltage-gated calcium channels,'' \emph{Cold Spring Harb.
  Perspect. Biol.}, vol.~3, no.~8, p. a003947, 2011.

\bibitem{dibak2017msm}
M.~Dibak, M.~J. del Razo, D.~De~Sancho, C.~Sch{\"u}tte, and F.~No{\'e},
  ``{MSM}/{RD}: Coupling {M}arkov state models of molecular kinetics with
  reaction-diffusion simulations,'' \emph{J. Chem. Phys.}, vol. 148, no.~21, p.
  214107, 2018.

\bibitem{bowman2014introduction}
G.~R. Bowman, V.~S. Pande, and F.~No{\'e}, ``Introduction and overview of this
  book,'' in \emph{An introduction to {M}arkov state models and their
  application to long timescale molecular simulation}.\hskip 1em plus 0.5em
  minus 0.4em\relax Springer, 2014, pp. 1--6.

\bibitem{prinz2011markov}
J.-H. Prinz, H.~Wu, M.~Sarich, B.~Keller, M.~Senne, M.~Held, J.~D. Chodera,
  C.~Sch{\"u}tte, and F.~No{\'e}, ``Markov models of molecular kinetics:
  Generation and validation,'' \emph{J. chem. phys.}, vol. 134, no.~17, p.
  174105, 2011.

\bibitem{schutte2013metastability}
C.~Sch{\"u}tte and M.~Sarich, \emph{Metastability and {M}arkov state models in
  molecular dynamics: modeling, analysis, algorithmic approaches}.\hskip 1em
  plus 0.5em minus 0.4em\relax American Mathematical Soc., 2013, vol.~24.

\bibitem{trendelkamp2015estimation}
B.~Trendelkamp-Schroer, H.~Wu, F.~Paul, and F.~No{\'e}, ``Estimation and
  uncertainty of reversible {M}arkov models,'' \emph{J. chem. phys.}, vol. 143,
  no.~17, p. 11B601\_1, 2015.

\bibitem{sadeghi2018particle}
M.~Sadeghi, T.~R. Weikl, and F.~No{\'e}, ``Particle-based membrane model for
  mesoscopic simulation of cellular dynamics,'' \emph{J. Chem. Phys.}, vol.
  148, no.~4, p. 044901, 2018.

\bibitem{lin1991divergence}
J.~Lin, ``Divergence measures based on the {S}hannon entropy,'' \emph{{IEEE}
  Trans. Inf. Theory}, vol.~37, no.~1, pp. 145--151, 1991.

\end{thebibliography}

\renewcommand{\thesection}{\Alph{section}} \numberwithin{equation}{section}

\appendix

\section{Smoluchowski's model with periodic flux}

\label{sec:smol-per} In addition to the models of Sec. \ref{sec:modelsdiffinf},
we can also obtain a concentration-based Smoluchowski model in the
canonical ensemble, i.e. where the total concentration of $B$ is
conserved. This is achieved by using a partially absorbing boundary
condition and forcing a periodic flux. The corresponding boundary
conditions for the Fokker-Planck equation (Eq. (\ref{eq:smol})) are,
\begin{align}
\left.4\pi\sigma^{2}D\frac{\partial f(r,t)}{\partial r}\right|_{r=\sigma}=\left.4\pi R^{2}D\frac{\partial f(r,t)}{\partial r}\right|_{r=R}=\kappa f(\sigma,t),\label{CKperiodic}
\end{align}
and $\int_{\sigma}^{R}4\pi r^{2}f(r,t)dr=1$. These conditions mean
that the probability flux at $r=\sigma$ is the same as the flux at
$r=R$. The steady state solution is exactly of the same form as Eq.
(\ref{eq:cksol}), but with the constant $A_{0}$ instead of $c_{0}$,
\begin{align}
f^{ss}(r) & =A_{0}\left[1-\frac{\kappa\sigma}{4\pi D\sigma+\kappa}\left(\frac{1}{r}\right)\right],\label{eq:ck}\\
 & A_{0}=\left[4\pi\left(\frac{R^{3}-\sigma^{3}}{3}\right)-\frac{4\pi\sigma\kappa}{4\pi\sigma D+\kappa}\left(\frac{R^{2}-\sigma^{2}}{2}\right)\right]^{-1}.\nonumber 
\end{align}
This result is a first step to provide a mathematical connection between
theconcentration-based approach and the probability approach. It can
be easily interpreted as a concentration gradient for a large number
of $B$ molecules, where the absorbtion flux at $\sigma$ is exactly
the same as the incoming flux of particle at $r=R$, but it can also
be understood as the probability distribution for one $B$ molecule,
which every time it is absorbed at $r=\sigma$, it is placed back
again at $r=R$. 

We should note that the boundary condition in Eq. (\ref{CKperiodic})
is also satisfied in the original Collins and Kimball formulation
at steady state from Eq. (\ref{eq:cksol}). However, in a probabilistic
interpretation, the free parameter $A_{0}$ would give the normalization
constant for the probability, which we can fix so the probability
integrates to one. 

\section{Canonical Smoluchowski master equation\label{sec:canon}}

The SME derived in Sec. \ref{sec:SME} gives us the dynamics of the
probability of one $B$ molecule in this system. The quantity $\pi_{i}(t)$
is the probability of finding one $B$ molecule in shell $i$ at time
$t$. In this section, we will obtain the SME in the canonical ensemble
for an arbitrarily but fixed number of $B$ molecules. We assume $m$
independent and identical $B$ molecules that obey Eq. (\ref{eq:SME}).
The number of ways to arrange $m$ independent $B$ molecules in the
system, such that $n_{i}$ are in state $i$ (shell $i$) while maintaining
the total number constant $m=n_{1}+n_{2}+\dotsc+n_{N}$, is given
by the multinomial distribution \citep{heuett2006grand}. Therefore,
we can write the joint probability of having $n_{i}$ molecules on
each state simply as the multinomial 
\begin{equation}
P({\scriptstyle n_{0},n_{1},\dotsc,n_{N},t})=\frac{m!}{n_{0}!n_{1}!\dotsc n_{N}!}\pi_{0}(t)^{n_{0}}\pi_{1}(t)^{n_{1}}\dotsc\pi_{N}(t)^{n_{N}}.\label{eq:multi-Smol}
\end{equation}
Therefore, the expected value of having $n_{k}$ molecules in shell
$k$ at time $t$ is given by the expected value of the multinomial,
\begin{equation}
\mathbb{E}[\mathcal{N}_{k}=n_{k}]=m\pi_{k}(t),\label{eq:ev_mult}
\end{equation}
where $\mathcal{N}_{k}$ refers to the random number of particles
in shell $k$. We now show that the equation satisfied by this expected
value in the continuous limit is the Smoluchowski's equation (Eq.
(\ref{eq:smol})) with the periodic boundary conditions from Eq. (\ref{CKperiodic}).

In the interest of minimizing notation, we will refer to the expected
value of Eq. (\ref{eq:ev_mult}), $\mathbb{E}[\mathcal{N}_{k}=n_{k}]$,
as $F_{i}(t)=m\pi_{k}(t)$. We want to establish a connection between
this model and the original SME from Eq. (\ref{eq:SME}). In order
to do so, we only need to multiply by $m$ the equation for the $i^{th}$
shell given by Eq. (\ref{eq:SME-ind}). This yields 
\begin{equation}
\frac{dF_{i}(t)}{dt}=q_{i+1,i}F_{i+1}(t)-(q_{i,i-1}+q_{1,i+1})F_{i}(t)+q_{i-1,i}F_{i-1}(t).\label{eq:FD-expval}
\end{equation}

We will now follow a similar procedure to that of \citep{del2016discrete}.
Substituting the corresponding values for the transition rates given
in Eq. (\ref{probdiff2}), we obtain the following equation

{\small{}
\begin{align}
\frac{dF_{i}(t)}{dt}=D\left[\frac{F_{i+1}(t)-2F_{i}(t)+F_{i-1}(t)}{\delta r^{2}}\right]\nonumber \\
-\frac{2D}{r_{i}}\left[\frac{F_{i+1}(t)-F_{i-1}(t)}{2\delta r}\right]+\frac{D}{\delta r}\left[\frac{F_{i}(t)}{r_{i}-\delta r}-\frac{F_{i}(t)}{r_{i}+\delta r}\right].\label{eq:ithSME}
\end{align}
} We can now take th limit as $\delta r\to0$ to obtain 
\begin{align}
\frac{\partial F(r,t)}{\partial t} & =D\frac{\partial^{2}F(r,t)}{\partial r^{2}}-\frac{2D}{r}\frac{\partial F(r,t)}{\partial r}+\frac{2D}{r^{2}}F(r,t),\nonumber \\[3mm]
 & =D\frac{\partial^{2}F(r,t)}{\partial r^{2}}-\frac{\partial}{\partial r}\left(\frac{2D}{r}F(r,t)\right),\label{eq:FPE-dr}
\end{align}
where $F(r,t)dr$ is the continuous analog of $F_{i}(t)$, i.e. the
expected value for the number of $B$ particles in a shell of width
$\delta r$ in position $r$ and at time $t$.

This is the expected value computed at any point in the shell with
radius $r$, so we cannot yet compare it with the Smoluchowski diffusion
equation. In order to do so, we need the equation for the expected
value at any point in space given by $f(r,\theta,\phi,t)r^{2}\sin(\theta)drd\theta d\phi$.
Integrating this equation in the angular coordinates due to symmetry
yields the expected value we just obtained, $F(r,t)$, 
\begin{equation}
F(r,t)dr=4\pi r^{2}f(r,t)dr.\label{eq:spscaling}
\end{equation}
Substituting this result into Eq. (\ref{eq:FPE-dr}) and doing some
algebra, we recover the Smoluchowski original equation, Eq. (\ref{eq:smol}),
\[
\frac{\partial f(r,t)}{\partial t}=\frac{D}{r^{2}}\frac{\partial}{\partial r}\left(r^{2}\frac{\partial f(r,t)}{\partial r}\right).
\]
Note that, in this case, the equation has a very precise meaning.
The quantity $4\pi r^{2}f(r,t)\delta r$ is the
expected number of particles at the shell of radius $r$ and width
$\delta r$ at time $t$. More precisely, the quantity $f(r,t)$ has
units of number of particles per unit volume, so it is the expected
value for the concentration at a given point with position $r$ at
time $t$.

We still need to deal with the boundary conditions. We can also obtain
the equations at the boundaries by again multiplying by $m$ the first
and last equation of the system of Eqs. (\ref{eq:SME}). The resulting
equations for the inner and outer boundaries are the following, 
\begin{align}
\frac{dF_{0}(t)}{dt} & =-(q_{0,1}+q_{0,b})F_{0}(t)+q_{1,0}F_{1}(t)\label{eq:FD-innerBC}\\
\frac{dF_{N}(t)}{dt} & =F_{0}(t)q_{0,b}+F_{N-1}(t)q_{N-1,N}-F_{N}(t)q_{N,N-1}\nonumber 
\end{align}
Note $q_{0,b}=\tilde{\kappa}(r)$, where the physically reasonable
assumption is that the rate $\tilde{\kappa}(r)$ scales inversely
to the infinitesimal volume of the reaction spherical shell, i.e.
$\tilde{\kappa}(r)=\kappa/(4\pi r^{2}\delta_{r})$, where $\kappa$
will be the constant rate in the boundary condition \citep{del2016discrete}.

Substituting the rates into the last two equations at the inner and
outer boundary at shells $i=0$ and $i=n$ and doing some algebra,
we obtain the following equations for the inner boundary , 
\begin{align}
\frac{dF_{0}}{dt}=\frac{D}{\delta r^{2}}\left[F_{1}-2F_{0}+F_{0}\left(1-\frac{\delta r^{2}}{D}\frac{\kappa}{4\pi r_{0}^{2}\delta r}\right)\right]\nonumber \\[2mm]
-\left[\frac{D}{r_{0}}\frac{F_{1}}{\delta r}+\frac{D}{\delta r}\left(\frac{F_{0}}{r_{0}+\delta r}\right)\right]\nonumber \\
=D\left[\frac{F_{1}-2F_{0}+F_{-1}}{\delta r^{2}}\right]-\frac{2D}{r_{0}}\left[\frac{F_{1}-F_{-1}}{2\delta r}\right]\nonumber \\
+\frac{D}{\delta r}\left[\frac{F_{0}}{r_{0}-\delta r}-\frac{F_{0}}{r_{0}+\delta r}\right],\label{eq:meBCin}
\end{align}
and for the outer boundary 
\begin{align}
\frac{dF_{N}}{dt} & =\frac{D}{\delta r^{2}}\left[F_{N-1}-2F_{N}+F_{N}\right]\nonumber \\[2mm]
 & +\left[\frac{D}{r_{N}}\frac{F_{N-1}}{\delta r}+\frac{D}{\delta r}\left(\frac{F_{N}}{r_{N}-\delta r}\right)\right]-F_{0}\frac{\kappa}{4\pi r_{0}^{2}\delta r},\nonumber \\
 & =D\left[\frac{F_{N-1}-2F_{N}+F_{N+1}}{\delta r^{2}}\right]-\frac{2D}{r_{N}}\left[\frac{F_{N+1}-F_{N-1}}{2\delta r}\right]\nonumber \\
 & +\frac{D}{\delta r}\left[\frac{F_{N}}{r_{N}-\delta r}-\frac{F_{0}}{r_{N}+\delta r}\right].\label{eq:meBCout}
\end{align}
Note we omitted the time dependence of $F_{i}(t)$ to simplify notation.
In both cases, Eq. (\ref{eq:meBCin}) and Eq. (\ref{eq:meBCout}),
we introduced the ghost cells $F_{-1}$ and $F_{N+1}$ respectively
to force the equation to satisfy Eq. (\ref{eq:ithSME}) (the equation
satisfied inside the boundaries). In order for Eqs. (\ref{eq:meBCin})
and (\ref{eq:meBCout}) to be satisfied, the ghost cells need to satisfy
the equations 
\begin{align*}
F_{0}-F_{0}\frac{\kappa\delta r}{4\pi Dr_{0}^{2}} & =F_{-1}+\frac{\delta r}{r_{0}}F_{-1}+\frac{\delta r}{r_{0}-\delta r}F_{0},\\
F_{N}-F_{0}\frac{\kappa\delta r}{4\pi Dr_{0}^{2}} & =F_{N+1}-\frac{\delta r}{r_{N}}F_{N+1}-\frac{\delta r}{r_{N}+\delta r}F_{N},
\end{align*}
which will yield the boundary conditions. Arranging terms, dividing
by $\delta r$ and taking the limit as $\delta r\to0$, we obtain
\begin{align*}
\left.\frac{\partial F(r,t)}{\partial r}\right|_{r=\sigma} & =\frac{\kappa}{4\pi D\sigma^{2}}F(\sigma,t)+\frac{F(\sigma,t)}{\sigma},\\[2mm]
\left.\frac{\partial F(r,t)}{\partial r}\right|_{r=r_{\mathrm{max}}} & =\frac{\kappa}{4\pi D\sigma^{2}}F(\sigma,t)+\frac{F(r_{\mathrm{max}},t)}{r_{\mathrm{max}}},
\end{align*}
respectively, where $r_{0}=\sigma$ is the innermost shell and $r_{\mathrm{max}}$
is the outermost shell. Applying once again the identity in Eq. (\ref{eq:spscaling}),
we obtain the boundary conditions for the Smoluchowski model with
periodic flux from Appendix \ref{sec:smol-per}, 
\begin{align*}
\left.4\pi D\sigma^{2}\frac{\partial f(r,t)}{\partial r}\right|_{r=\sigma}=\kappa f(\sigma,t),\\[2mm]
\left.4\pi Dr_{\mathrm{max}}^{2}\frac{\partial f(r,t)}{\partial r}\right|_{r=r_{\mathrm{max}}}=\kappa f(\sigma,t).
\end{align*}
These are the boundary conditions for the expected value of the concentration
at position $r$ and time $t$. It should be noted that the process
to obtain the continuous limit of these equations is analogous to
the one we presented in \citep{del2016discrete}.

This result shows that the Smoluchowski model with periodic flux from
Appendix \ref{sec:smol-per} is the mean field of Eq. (\ref{eq:multi-Smol}), i.e. 
a number $m$ of $B$
molecules, each obeying Eq. (\ref{eq:SME}). The model described
by Eq. (\ref{eq:multi-Smol}) provides a probabilistic approach to
model concentration-based diffusion-influenced reactions. It not only
yields the expected mean field but can also yield the full probability
distribution for all particles.

The steady state of the probabilistic model from Eq. (\ref{eq:multi-Smol})
is a nonequilibrium steady state since it always has a constant flux
from the outer boundary into the inner one, so the total number of
particles in the system does not change or fluctuate over time. Therefore,
following statistical mechanics terminology, we say this system is
in the canonical ensemble. Note the original Smoluchowski concentration-based
model (Eq. (\ref{eq:smol})) does not maintain a constant number of
particles (or concentration), unless the system is in steady state.
Therefore, the model from Eq. (\ref{eq:multi-Smol}) has limited applicability.
Nonetheless, this is the first step to connect the probabilistic and
concentration interpretations. For a general case, see Sec. \ref{sec:grand}.

\section{Large copy number limit \label{sec:largecopylim}}

In this section, we show the large copy number limit
of the GC-SME also recovers the concentration-based approach from
Sec. \ref{sec:difreac}. We begin by pointing out that Eq. (\ref{eq:FD-expval2})
is a particular case of the equation, 
\[
\frac{d\left\langle n_{i}(t)\right\rangle }{dt}=\sum_{\substack{j=-1\\
j\ne i
}
}^{N+1}\left[\left\langle n_{j}(t)\right\rangle q_{j,i}-\left\langle n_{i}(t)\right\rangle q_{i,j}\right].
\]
which corresponds to a generalized version of the master equation,
Eq. (\ref{eq:GCE-mastereq}), where all the states can interact with
one another. In \citep{heuett2006grand}, it was shown that a solution
to this general master equation satisfies the following Poisson probability
distribution, 
\begin{equation}
P(n_{0},n_{1},\dotsc,n_{N},t)=\prod_{i=0}^{N}\left[\frac{\left\langle n_{i}(t)\right\rangle {}^{n_{i}}}{n_{i}!}e^{-\left\langle n_{i}(t)\right\rangle }\right].\label{eq:p_poisson}
\end{equation}
This can be proved by direct substitution. As our equation is of the
same form, it also satisfies the same distribution. Also note the
expected value for the number of particles at shell $i$ of this distribution
is $\left\langle n_{i}(t)\right\rangle $. Taking the marginal distribution
by integrating all except one of the variables, we obtain 
\begin{equation}
P(\mathcal{N}_{j}=n_{j},t)=\frac{\left\langle n_{j}(t)\right\rangle ^{n_{j}}}{n_{j}!}e^{-\left\langle n_{j}(t)\right\rangle }.\label{eq:marg_poisson}
\end{equation}
The random variable $\mathcal{N}_{j}$ which gives the number of particles
at each state/shell obeys a simple Poisson distribution. Now consider
the scaling $\mathcal{N}_{j}^{\text{fr}}=\mathcal{N}_{j}/c_{0}$ ($c_{0}$
constant). The mean and standard deviation for $\mathcal{N}_{j}^{\text{fr}}$
are then given by 
\[
\mu_{j}^{\text{fr}}=\frac{\left\langle n_{j}(t)\right\rangle }{c_{0}}\ \ \ \ \ \sigma_{j}^{\text{fr}}=\frac{\left\langle n_{j}(t)\right\rangle }{c_{0}}
\]
Using Chebyshev's inequality, we obtain 
\begin{align}
\Pr(|\mathcal{N}_{j}^{\text{fr}}-\mu_{j}^{\text{fr}}|\le\epsilon)\ge1-\frac{\left(\sigma_{j}^{\text{fr}}\right)^{2}}{\epsilon^{2}}=1-\frac{\left\langle n_{j}(t)\right\rangle }{\epsilon^{2}c_{0}^{2}}\nonumber \\[2mm]
\Rightarrow\ \ \ \ \ \ \ \Pr(|\mathcal{N}_{j}^{\text{fr}}-\mu_{j}^{\text{fr}}|>\epsilon)<\frac{\left\langle n_{j}(t)\right\rangle }{\epsilon^{2}c_{0}^{2}}.\label{eq:GC_lln}
\end{align}
As the total number of particles in the bath goes to infinity ($c_{0}\rightarrow\infty$),
the average number of particles $\left\langle n_{j}(t)\right\rangle $
at any given shell will also go to infinity such that the ratio $\mu_{j}^{\text{fr}}=\left\langle n_{j}(t)\right\rangle /c_{0}$
is fixed. We know this ratio is finite because the equation satisfied
by $\mu_{j}^{\text{fr}}$ is nothing more than a scaled version of
Eq. (\ref{eq:FD-expval2}). Nonetheless, the ratio $\left\langle n_{j}(t)\right\rangle /c_{0}^{2}$
will go to zero as $c_{0}\rightarrow\infty$; therefore, this inequality
implies $\mathcal{N}_{j}^{\text{fr}}$ will approach $\mu_{j}^{\text{fr}}$
in the large copy number limit, i.e. the law of large numbers. Consequently,
$\mathcal{N}_{j}^{\text{fr}}$ will also approach the solution of
Eq. (\ref{eq:FD-expval2}) scaled by $c_{0}$. Furthermore, we can
carry the continuous limit for Eq. (\ref{eq:FD-expval2}) scaled by
$c_{0}$, where the solution is $f^{\text{fr}}(r,t)=f(r,t)/c_{0}$.
Consequently, the large particle number limit ($c_{0}\rightarrow\infty$)
of the ratio between local concentration and the bath state concentration
satisfies the Smoluchowski equation with $f^{\text{fr}}(r_{\text{max}},t)=1$. 

Note we could have normalized the random variable
$\mathcal{N}_{j}$ by any constant and obtain that $\int_{\sigma}^{r_{\text{max}}}f^{\text{fr}}(r,t)4\pi r^{2}dr\le1$,
which means that $f^{\text{fr}}4\pi r^{2}$ can be easily confused
with a probability distribution function; however, it is not.
\end{document}